  \documentclass[11pt,epsf,aps, article]{revtex4}{}     
\usepackage{graphicx}
\usepackage{latexsym,amsmath,amssymb,bm,euscript,eufrak}
\usepackage{hyperref}

\begin{document}

\title{Amperean pairing and the pseudogap phase of cuprate superconductors}
\author{Patrick A. Lee}
\email{palee@mit.edu}
\affiliation{Department of Physics, Massachusetts Institute of Technology, Cambridge, MA 02139}

\date{\today}

\begin{abstract}
The enigmatic pseudogap phase in underdoped cuprate high $\rm T_c$ superconductors has long been recognized as a central puzzle of the $\rm T_c$ problem.  Recent data show that the pseudogap is likely a distinct phase, characterized by a medium range and quasi-static charge ordering.  
However, the origin of the ordering wavevector and the mechanism of the charge order is unknown.
At the same time, earlier data show that precursive superconducting fluctuations are also associated with this phase.  
We propose that the pseudogap phase is a novel pairing state where electrons on the same side of the Fermi surface are paired, in strong contrast with conventional BCS theory which pair electrons on opposite sides of the Fermi surface.  In this state the Cooper pair carries a net momentum and belong to a general class called pair density wave (PDW).  The microscopic pairing mechanism comes from a gauge theory formulation of the resonating valence bond (RVB) picture, where spinons traveling in the same direction feel an attractive force in analogy with Ampere's effects in electromagnetism.  We call this Amperean pairing.  Charge order automatically appears as a subsidiary order parameter even when long range pair order is destroyed by phase fluctuations.  Our theory gives a prediction of the ordering wavevector which is in good agreement with experiment.  Furthermore, the quasiparticle spectrum from our model explains many of the unusual features reported in photoemission experiments.  The Fermi arc, the unusual way the tip of the arc terminates and the relation of the spanning vector of the arc tips to the charge ordering wavevector also come out naturally.  We also discuss how the onset of the Kerr effect in this state can be accommodated.  Finally, we propose an experiment which can directly test the notion of Amperean pairing.
\end{abstract}

\pacs{74, 73}
\maketitle

Since the early days of cuprate superconductivity research, the pseudogap phase has been identified as a central piece of the high {\rm T$_c$} puzzle.\cite{1} The pseudogap opens below a temperature $T^\ast$ much above {\rm T$_c$} in underdoped cuprates, and is visible in the spin susceptibility as detected by the Knight shift, in tunneling spectroscopy and in {\em c}-axis conductivity.  Angle-resolved photoemission (ARPES) shows that the pseudogap opens in the anti-nodal region near $(0,\pi)$, (we set the lattice constant $a$ to unity),  leaving behind ungapped ``Fermi arcs'' centered around the nodes.  Recent X-ray scattering data \cite{2,3,4,5} reveal that the pseudogap is likely to be a distinct phase, characterized by the onset of a charge density wave (CDW) with wavevectors at $(0,\pm \delta)$ and $(\pm \delta,0)$ where $\delta$ {\em decreases} with increasing doping, thus confirming evidence for charge order found   earlier by  STM  \cite{5a,5b,5c,6} and NMR experiments. \cite{6a} Recent advances include STM and X-ray studies on the same Bi-2212 samples. \cite{6b} Other signatures include Kerr rotation,\cite{7} the emergence of anisotropy in the Nernst effect, etc.\cite{8}  (Another set of experiments found the onset of intra unit cell magnetization (loop currents) at a somewhat higher temperature.\cite{9}  We shall not address this phenomenon in this paper.)

The CDW is enhanced by a magnetic field and appears to be 
connected to the high magnetic field state where quantum oscillations have been observed.\cite{10}  Indeed, recent work \cite{11} claims that the intrinsic $H_{c2}$ of some underdoped YBCO samples may be as low as 22~T, a shockingly low energy scale compared with the energy gap scale as well as {\rm$T_c$}.  At the same time the onset of superconductivity causes a reduction of the CDW amplitude,\cite{2,3,4} suggesting that the pseudogap should be considered a competing phase.  However, up to now the identity of this phase is not at all clear because the data seem to be giving conflicting signatures. Certain features of the pseudogap state suggests the presence of short range superconducting order. For example, diamagnetic fluctuations are observed much above {\rm T$_c$} \cite{12} and 
the spectral weight of the Drude conductivity between members of the bilayers in YBCO is  found to increase below $\sim$180~K, a trend consistent with fluctuating superconductivity and opposite to gap formation due to charge order.
\cite{13} On the other hand, the CDW is rather long ranged in the plane, but it does not resemble a conventional CDW driven by Fermi surface nesting.  A recent comparison of $\delta$ with the nesting vector connecting the Fermi surfaces at the anti-nodal points in single layer Bi$_2$Sr$_{2-x}$La$_x$CuO$_{6+\delta}$ (Bi2201) found that $\delta$ is larger by about a factor of 2. Instead the CDW vector was suggested to connect the ends of the Fermi arcs where the gap is zero or very small.\cite{5} This directly contradicts the standard CDW picture where an energy gap is expected at precisely that point (see supplementary material).  As we discuss in greater detail later, this difficulty is anticipated in an earlier study where detailed ARPES data show that the spectrum contains highly anomalous features which cannot be understood with a CDW model.\cite{7}  In this paper we propose a new model for the pseudogap state which combines pairing and CDW and addresses all the mysterious properties described above.

Our model is based on the idea of Amperean pairing introduced earlier in the context of quantum spin liquids.\cite{14} The U(1) spin liquid is described by a spinon Fermi surface coupled to a U(1) gauge field.  In this system the coupling to transverse gauge field (gauge magnetic field) is strong and unscreened.  Just as in electrodynamics, the current of a carrier produces a gauge magnetic field which creates an attractive force to another carrier moving in the same direction, due to the Ampere effect.  The transverse gauge field mediated interaction contributes the following to the action,\cite{14}
\begin{equation}
S_{int} = -\frac{1}{2v\beta} \sum_{{\bm p}_1{\bm p}_2 , {\bm q}} D(q) 
(\bm v_{\bm p_1} \times \hat{q}) \cdot (\bm v_{\bm p_2} \times \hat{q})
f^\dagger_{\bm p_1+\bm q,\sigma} f^\dagger_{\bm p_2-\bm q,\sigma^\prime} f_{\bm p_2,\sigma^\prime} f_{\bm p_1,\sigma}
\end{equation}
where $v$ is the volume, $\beta$ = 1/kT, $\bm{v_p}$ is the velocity of the spinon with momentum $\bm p$, $D(\bm q)$ is the gauge field propagator and $f_{\bm p, \sigma}$ is the spinon destruction operator with spin $\sigma$.  In ref. \cite{14} we showed that due to the singular nature of the gauge propagator, the Fermi surface is unstable to a special kind of pairing where particles in the vicinity of a given spot $\bm K$ on the Fermi surface form Cooper pairs.  This is radically different from conventional BCS pairing, which pair particles on opposite sides of the Fermi surface.  We called this Amperean pairing.

In one version of the RVB theory of high $\rm T_c$ superconductors, a spinon Fermi surface is formed which is coupled to a U(1) gauge field.\cite{1} Unlike the spin liquid, there is also a gapless charge degree of freedom which is described by a bosonic holon coupled to the same gauge field.  The bosons tend to condense and convert the spinons to electrons with a reduced spectral weight equal to $p$, the hole doping concentration.  While the gauge field is gapped by the Anderson-Higgs mechanism, the gap is small for small $p$ and gauge fluctuations remain important over a large temperature range, which is referred to as the incoherent Fermi liquid region.\cite{15}  Since it is in this region that the pseudogap state is formed, it is reasonable to assume that the same Amperean mechanism is at work and Eq.(1) still applies. Up to now the standard RVB picture is that d wave pairing of spinons onsets below a certain temperature.\cite{16} Instead, we assume that the Amperean pairing has a slightly lower free energy and pre-empts the d wave pairing.
This is reasonable because many states are competitive in energy with the d-wave state, including pair density wave states which share common properties with Amperean pairing as discussed below.\cite{21a}  As we shall see, Amperean pairing leaves segments of gapless excitations which contribute to a T linear entropy, exceeding the T$^2$ term for d-wave pairing.  Consequently, even if d-wave pairing is the true ground state, the Amperean state can have a lower free energy above some temperature.  
 In the cuprate it is natural to view the  hole Fermi surface as the analog of the large and almost circular Fermi surface in the spin liquid problem.  As shown in Fig.1, 
\begin{figure}[h,b]
\includegraphics[width=4.25in]{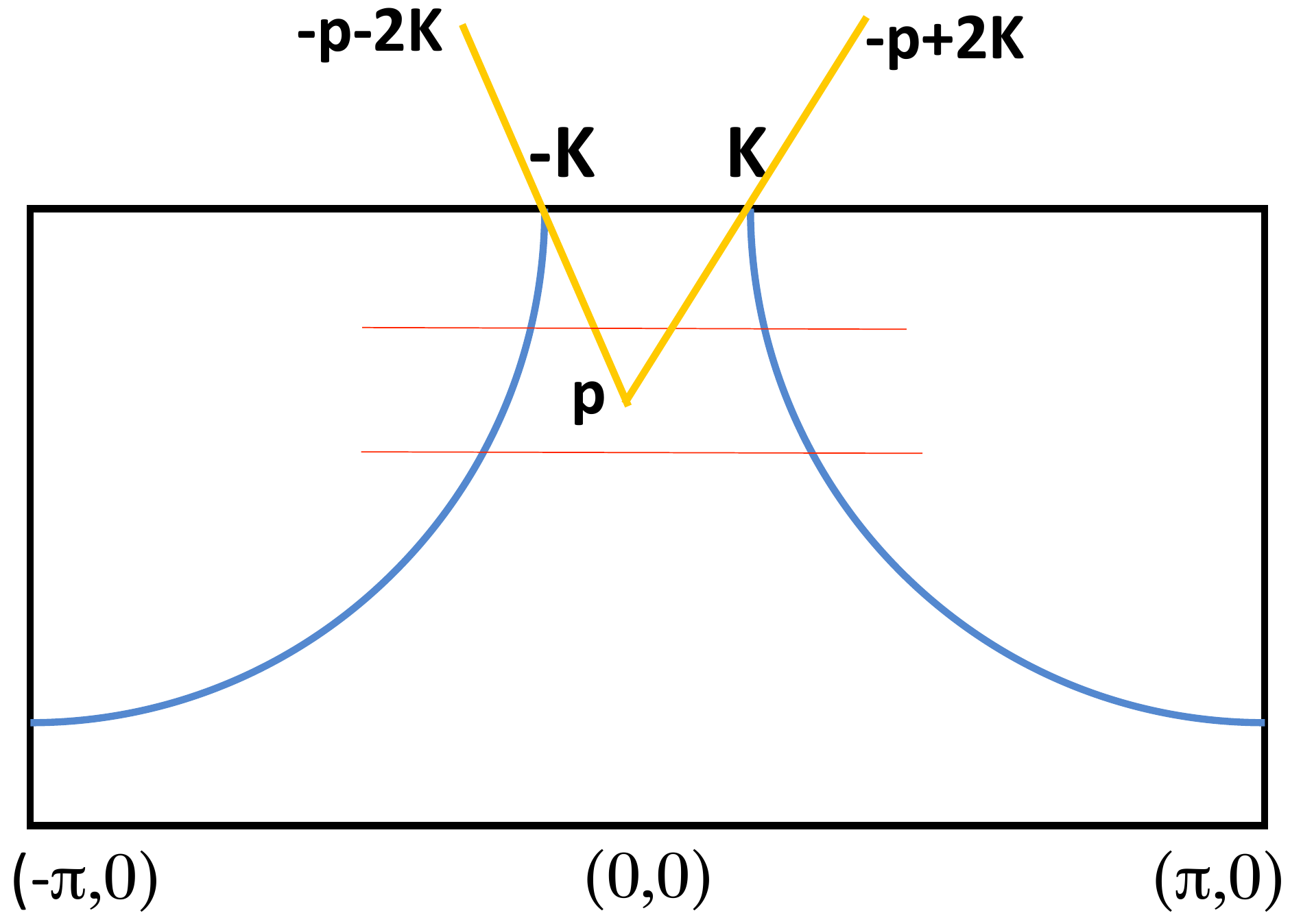}
\caption{Schematic representation of the top half of the cuprate Brillouin zone and the Fermi surface.  Pairs are formed between fermions near ${\bf K}$ with momenta ${\bf p}$ and ${-\bf p} + 2{\bf K}$ and similarly near $-2{\bf K}$.  The Cooper pairs carry momentum ${\bf Q} = 2{\bf K}$ and $-2{\bf K}$, respectively.  Red lines mark some of the scans shown in Fig.3.
}
\end{figure}
we choose the points $\bm K$ and $-\bm K$ (up to a reciprocal lattice vector) on the Fermi surface at the antinode as the points of Amperean pairing.  It is convenient to measure all momenta from $(0,\pi)$.  Let us introduce the mean field decoupling of Eq.(1)
\begin{equation}
S^{\rm MF} =  \Delta_{2\bm K}^\ast (\bm k) f_{\bm k\uparrow} f_{-\bm k + 2{\bm K}\downarrow} +c.c. 
+ \Delta_{-2\bm K}^\ast (\bm k) f_{\bm k\uparrow} f_{-\bm k - 2\bm K \downarrow} + c.c.  
\end{equation}
where we have set $\sigma = \downarrow$, $\sigma^\prime = \uparrow$, ${\bm p}_2 = \bm k$, ${\bm p}_1 = -{\bm k} \pm 2{\bm K}$ in Eq.(1) and $\Delta_{\pm 2\bm K}({\bm k})$  comes from the remainder of the equation and  is proportional to a sum over $\bm k^\prime$ of $\left< f_{\bm k^\prime\uparrow} f_{-\bm k^\prime \pm 2\bm K\downarrow} \right>$.  Instead of attempting a self consistent solution of the mean field equation, in this paper we simply assume a reasonable form of  $\Delta_{\pm 2\bm K}(\bm k)$ and explore the resulting quasiparticle structure.  This is an essential first step because the quasiparticle spectrum is quite different from our intuition based on the convention BCS pairing.  Our ansatz is
\begin{equation}
\Delta_{\pm 2\bm K}(\bm k) = f(k_y) \Delta_0
\end{equation}
where $f(k_y) = e^{(\pi-k_y)^2/k_0^2}$ with $k_0$ chosen to represent the fact that  the pairing should be limited to the vicinity of $\bm K$. 

We see from Eq.(2) that the pairing order parameters carry a total momentum of $2\bm K$ or $-2\bm K$.\cite{21b}  This is because the pair is made up of two fermions moving in  the same  direction and the total momentum is $\pm 2\bm K$.  By choosing $\Delta_{2\bm K}$ and $\Delta_{-2 \bm K}$ to have the same amplitude, the order parameter is modulated in space as $\Delta (\bm r) \approx \cos (2\bm K \cdot \bm r)$.  This belongs to a general class of pairing which has been named pair density wave (PDW).\cite{16a,17} Historically the first example of PDW is the LOFF state (or more precisely the LO state) where the PDW is argued to be more stable than the uniform state when the Fermi surfaces are split by Zeeman splitting.\cite{18,19} More recently another example  has been introduced in connection with stripe formation in the LBCO system near doping $p =1/8$.\cite{17}  We shall refer to this as stripe PDW.  There the spins form a spin density wave with wavevector  $Q = 2\pi/8a$ which is interpreted as anti-phase Neel states separated by charged domain walls which produces charge order at wavevector $2Q= 2\pi/4a$. The superconductor is assumed to be modulated at the same $Q$ and interpreted as $d$ wave superconductors with anti-phase domains.  The state was introduced to explain the observation that while a superconducting state exists much above $\rm T_c$ within each layer, the layers fail to order coherently.\cite{20,21} Indeed, the stripe PDW has been shown to be energetically competitive in earlier projected wavefunction studies \cite{22}  and has been suggested as being stabilized in high magnetic fields.\cite{28a} We note that the microscopic picture of this state is very different from our Amperean pairing state.  The stripe-PDW begins with the stripe picture of the SDW and up to now has been discussed only in connection with the stripe phenomena near $p = 1/8$.  The Amperean pairing picture is not associated with any SDW order. Instead the wavevector is given in terms of Fermi surface spanning vectors which {\em decreases} with increasing doping, a trend opposite to that of stripe PDW.  The main driving force is the pairing energy which can be a high temperature scale. We may refer to our state as 2$k_F$ PDW.  On the other hand, in common with the stripe PDW, there is a CDW with period $2\bm Q = 4\bm K$ associated with Amperean pairing.  It is quite possible that  Amperean pairing help stabilize the stripe-PDW state in certain materials such as LBCO which favor SDW ordering and stripe formation. The stripe PDW is then separated  from the high temperature Amperean pair state by the phase transition near 50K. We emphasize that in our view, Amperean pairing is the driver and the CDW is a subsidiary order parameter.

Let us first consider $\bm k$ in the vicinity of $\bm K$ and calculate the spectrum due to the coupling of $c_{\bm k \uparrow}$ and $c_{-\bm k + 2\bm K \downarrow}^\dagger$ in the presence of $\Delta_{2\bm K}$.  Diagonalization of a $2 \times 2$ matrix gives
\begin{equation}
E_{k\uparrow}^\pm = \tfrac{1}{2} \left( \xi_{\bm k} - \xi _{-\bm k + \bm Q} \right) \pm
\sqrt{ \tfrac{1}{4} \left( \xi_{\bm k} + \xi _{-\bm k + \bm Q} \right) ^2 + | \Delta_{\bm Q} |^2}
\end{equation}
where $\xi_{\bm k} = \varepsilon_{\bm k} -\mu$ and $\bm Q = 2\bm K$. Eq.(4) replaces the familiar BCS spectrum  $\pm
\sqrt{ \xi_{\bm k}^2 + | \Delta_{\bm Q} |^2}$ and is no longer particle-hole symmetric. In Fig.2 we 
\begin{figure}[b]
\includegraphics[width=3.25in]{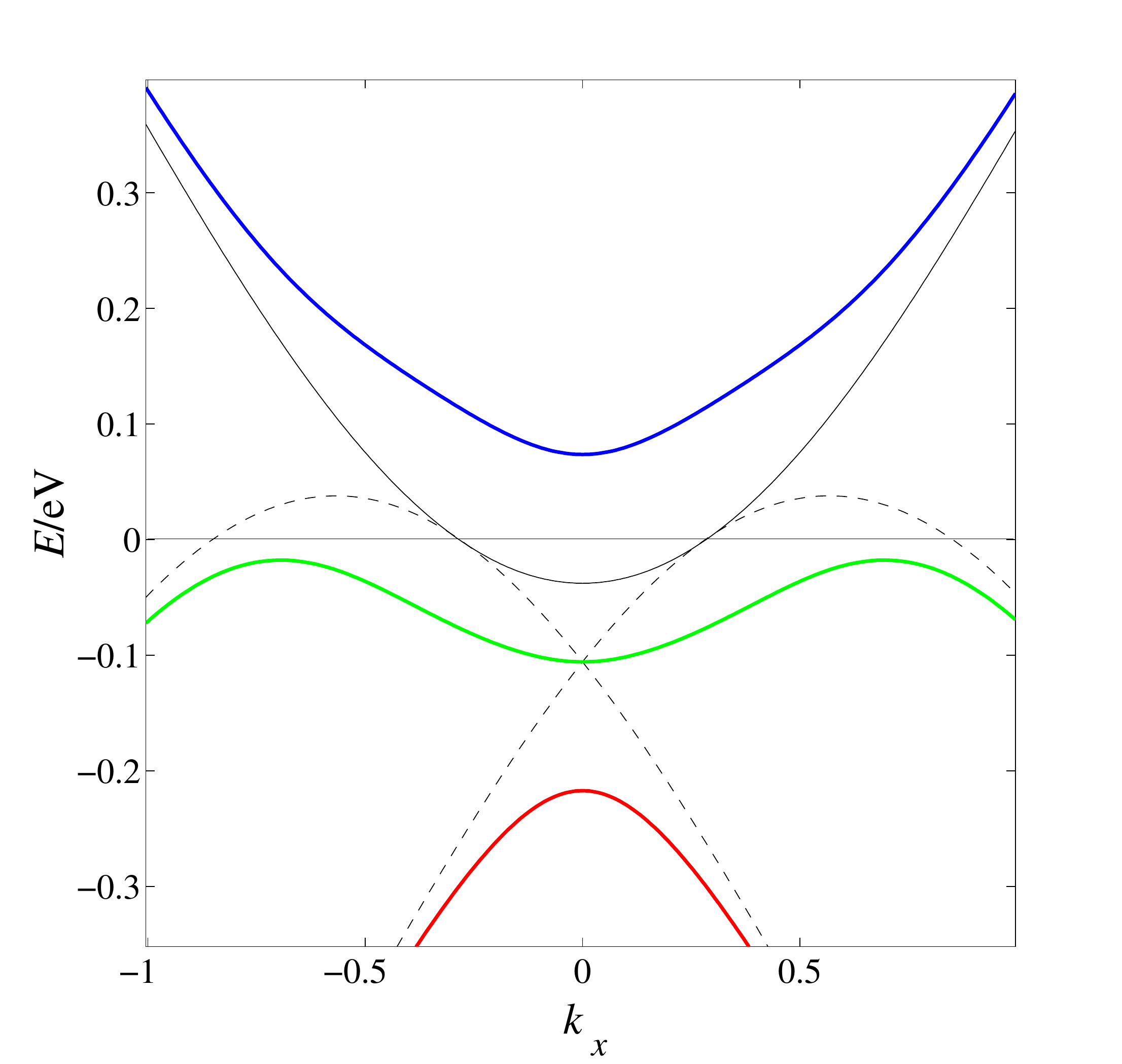}\includegraphics[width=3.25in]{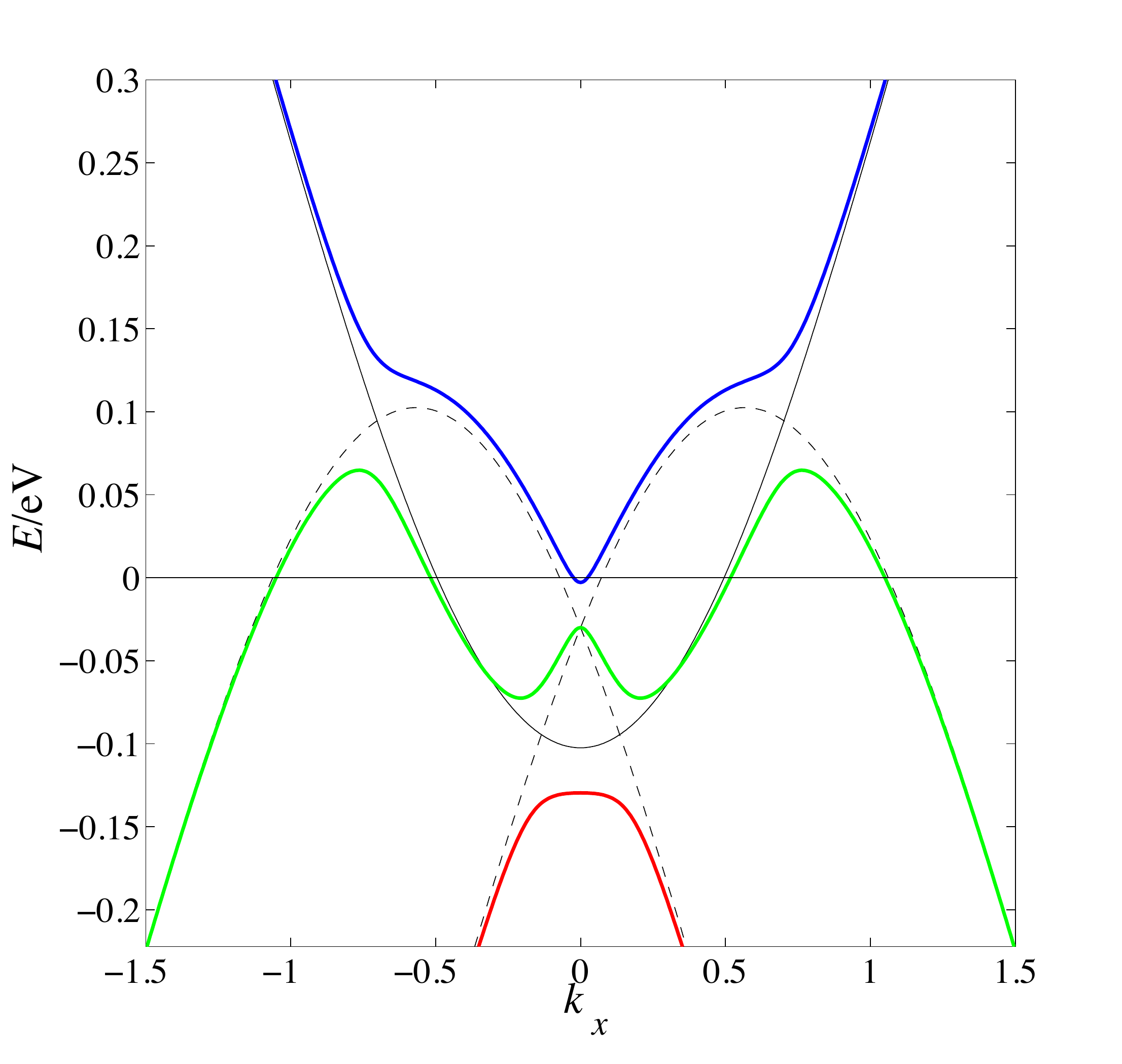}
\caption{Plot of the energy bands and the bare dispersion $\varepsilon_{\bm k}$ -$\mu$  versus  $k_x$ for (a) $k_y = \pi$ and (b) $k_y = \pi - 1.0$, using parameters appropriate for Bi-2201.]\cite{7}  The dashed lines shows $-(\varepsilon_{\bf k \pm 2{\bf K}}-\mu)$.  Turning on $\Delta_0 = 100$~meV with $k_0 = 1.0$ in Eq.(3) splits the bands and creates an energy gap for $k_y = \pi$.  The gap vanishes for $k_y = \pi -1.0$, creating a Fermi arc.}
\end{figure}
include in addition the coupling to $c_{-\bm k - 2\bm K \downarrow}^\dagger$ to restore inversion symmetry. We plot the band dispersions as well as the bare bands $\xi_{\bm k}$ and $-\xi_{-\bm k \pm  \bm Q}$ as a function of $k_x$ for $k_y = \pi$.\cite{28b}  Note that the bare bands touch each other at the Fermi level with parallel tangents.  Indeed if we ignore the band curvature and approximate $\xi_{\bm k} = v_F (k - k_F)$ in Eq.(2) , the two bands are exactly degenerate and are split by $| \Delta_{\bm Q} |$ for all $k_x$.  This gap opening is the source of energy gain of the Amperean pairing state.  Note that in this approximation, the Fermi surface remains gapless: the level splitting simply splits the Fermi vector.\cite{23}  However, once the band curvature is included and $|\Delta_{\bm Q}|$ is large enough, a gap is opened at the Fermi level near $(0,\pi)$, as shown in Fig. 2a.  As  $k_y$ moves away from $\pi/a$ the degeneracy is lifted and the effect of $\Delta_{\bm Q}$ diminishes, and eventually the Fermi surface is restored, as seen in Fig.2b.  This mechanism produces the Fermi arc.  
Importantly, the gap closes by occupied states arising up from below the Fermi energy, in contrast with a gap produced by CDW (see Appendix A).  We next consider the coupling between five states,  $\left(
c_{\bm k\uparrow}, c_{-{\bm k} + \bm Q \downarrow}^\dagger, c_{-\bm k - \bm Q \downarrow}^\dagger, c_{\bm k + 2\bm Q \uparrow}, c_{\bm k - 2\bm Q \downarrow}
\right)$  
and diagonalize the following $5 \times 5$ matrix .
\begin{equation}
\left[
\begin{array}{ccccc}
\xi_{\bm k} & \Delta_{\bm Q} & \Delta_{-\bm Q} & C_{2 \bm Q} & C_{-2 \bm Q}  \\
\\
\Delta^\ast_{\bm Q} & -\xi_{-\bm k + \bm Q} & 0 & 0 & \Delta^\ast_{-\bm Q}  \\
\\
\Delta^\ast_{\bm Q} & 0 & -\xi_{-\bm k-\bm Q} & \Delta^\ast_Q & 0 \\
\\
C^\ast_{2\bm Q} & 0 & \Delta_{\bm Q} & \xi_{\bm k + 2 \bm Q} & 0  \\
\\C^\ast_{-2 \bm Q} & \Delta_{-\bm Q} & 0 & 0 & \xi_{\bm k - 2\bm Q} 
\end{array}
\right]
\end{equation}
We added  $c_{\bm k \pm 2\bm Q,\uparrow}$ which are coupled to $c_{\bm k\uparrow}$ with matrix element $C_{\pm 2\bm Q}$ due to the CDW order  ,but more importantly they are also coupled to $c^\dagger_{-\bm k\pm \bm Q, \downarrow}$.  As seen from the full spectra in the supplementary material, the  latter bands are also degenerate at the Fermi level and give rise to splitting.  Inclusion of this coupling help create a pseudo-gap in the tunneling density of states which is closer to being particle-hole symmetric, as shown later.

In Fig.3 (f-i)
\begin{figure}[b]  
\includegraphics[width=6.5in]{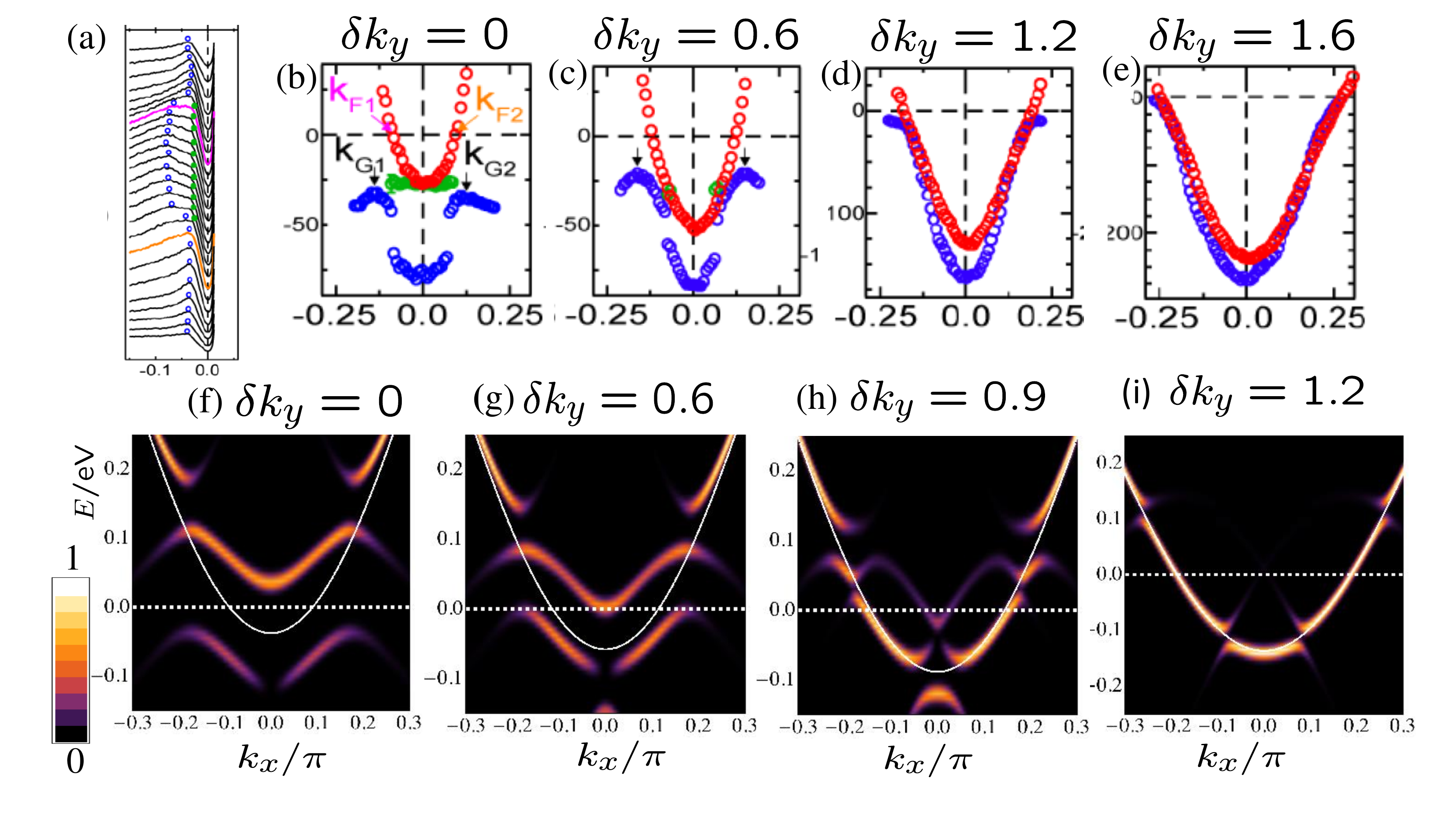}
\caption{ARPES spectra from ref.\cite{7} at 10~K ($<\rm T_c$) for scans with approximately constant $k_y = \pi -\delta k_y$ where $\delta k_y \approx 0, 0.6, 1.2$ and $1.6$.  Raw data for $\delta k_y = 0$ is shown in (a).  The Fermi arc begins near $\delta k_y = 1.2$.  The Amperean pairing spectra are shown $(f-i)$ for $\delta k_y = 0, 0.6, 0.9$ and $1.2$ using $\Delta_0 = 75$~meV and $k_0 =1$. The white line shows the band before pairing . With these parameters the gap closes and the Fermi arc begins near $\delta k_y = 0.6$. The negative energy states are to be compared with the data.  In (f), we note that the minimum excitation energy to the Fermi level does not lie at $k_F$ but at a larger wavevector.  This feature has been noted by the experimentalists, who introduced the wavevector $k_G$ in their figures (b) and (c) to describe this.  The green dots in (b) appear only below $\rm T_c$, and the blue dots at the band bottom mark the peak of a very broad spectrum near 100~meV as seen in (a).}
\end{figure}
we show a series of band dispersion with color intensity proportional to the spectral weight $| \nu_\alpha |^2$
where $\alpha = 1... 5$ labels the bands and $\nu_\alpha$ is the amplitude of the state $c_{\bm k\uparrow}$ for eigenstate $\alpha$.  The states below the Fermi level in this plot can be directly compared with ARPES data, which are reproduced in Fig.3 for Bi2201.  As discussed in ref.\cite{7}, the spectra show two highly unusual features.  While a gap is seen at the Fermi level for $k_y$ near $\pi$, the gap maximum is not at the Fermi momentum $k_F$ determined at high temperatures as expected for BCS pairing.  Instead the band extends beyond $k_F$ and  shows a peak at $k_G$ before losing spectral weight.  As seen in Fig. 3 (f-h) this unusual feature is captured by the Amperean pairing model.  As $k_y$ increases this effectively pushes the Fermi level crossing somewhat beyond the $k_F$ of the original band.  As shown in Fig. (4) ,this gives rise to a bending of the Fermi surface contour near the tip of the arc away from the high temperature Fermi surface, which is often seen in ARPES as well as STM data.\cite{24}  A more detailed examination of the spectrum in Appendix B shows that near the antinodes the top of the occupied band lies near $k_x=\pm Q$. If we define the tip of the Fermi arc as the $\bm k$ point where the top of the occupied band meets the Fermi level, this happens at $k_x=\pm Q$. Thus the vector connecting the tips of the arcs is the CDW wavevector, an observation made empirically by Comin et al \cite{5}. Due to the fact that the spectral weight is stronger for $|k_x|< Q$, the tip is rather ill-defined in Fig. 4 and tends to lie at $|k_x|$ slightly less than Q.  Finally, as seen in Fig. 3g, near the tip of the Fermi arc the band dispersion is not particle-hole symmetric, but simply turns around and loses spectral weight.  This feature has been emphasized by Yang {\em et al}.\cite{25} in their data, which they interpreted using a phenomenological model by Yang {\em et al}.\cite{31b}

A second unusual feature of the data of ref. \cite{7} is that near $k_y = \pi$, a broad spectral weight emerges around $-100$~meV, far below the band bottom at $- 35$~meV.  This feature persists to large $k_y$ and gives rise to the unusual feature that the spectrum appears broader at low temperature than above T$^\ast$.  In a conventional CDW model (see Appendix A) it is very hard to see where this spectral weight comes from.  In the Amperean pairing model, the band is connected to the crossing of the  $-\bm k \pm \bm Q$ hole bands located near 100 meV.   We expect these highly excited quasiparticles to be strongly coupled and the scattering between them can give rise to a broad line shape, as seen in experiment.  Physically these states should be interpreted as Andreev reflected hole states with momentum shifted by $\pm\bm Q$ due to the Amperean pairing amplitude $\Delta_{\pm \bm Q}$.

We see from Fig 3(f) that  the states at the saddle point $(k_x= 0)$ are pushed above the Fermi level. This is the origin of the weight seen near the $k_x$ and $k_y$ axes in Fig.4. When we couple to a conventional uniform BCS order parameter, these states will be split and produce coherent peaks in the usual way.  This is consistent with the feature in the data marked by the green dots.

In Fig.3 we have set the CDW coupling $C_{2\bm Q} = 0$.  It turns out that including a finite $C_{2\bm Q}$  does not significantly increase the energy gap and has little effect on the spectrum near the Fermi energy except that it opens a gap at the crossing of the vertical white line and the bare Fermi surface shown in Fig. 4.  This is because $C_{2\bm Q}$ couple the original bands far above the Fermi energy.  The observed pseudo gap is almost entirely due to Amperean pairing.  As discussed in Appendix A, it is not possible to explain the ARPES data in models where CDW alone is the driving force behind the energy gap.  

\begin{figure}[h]
\includegraphics[width=5.25in]{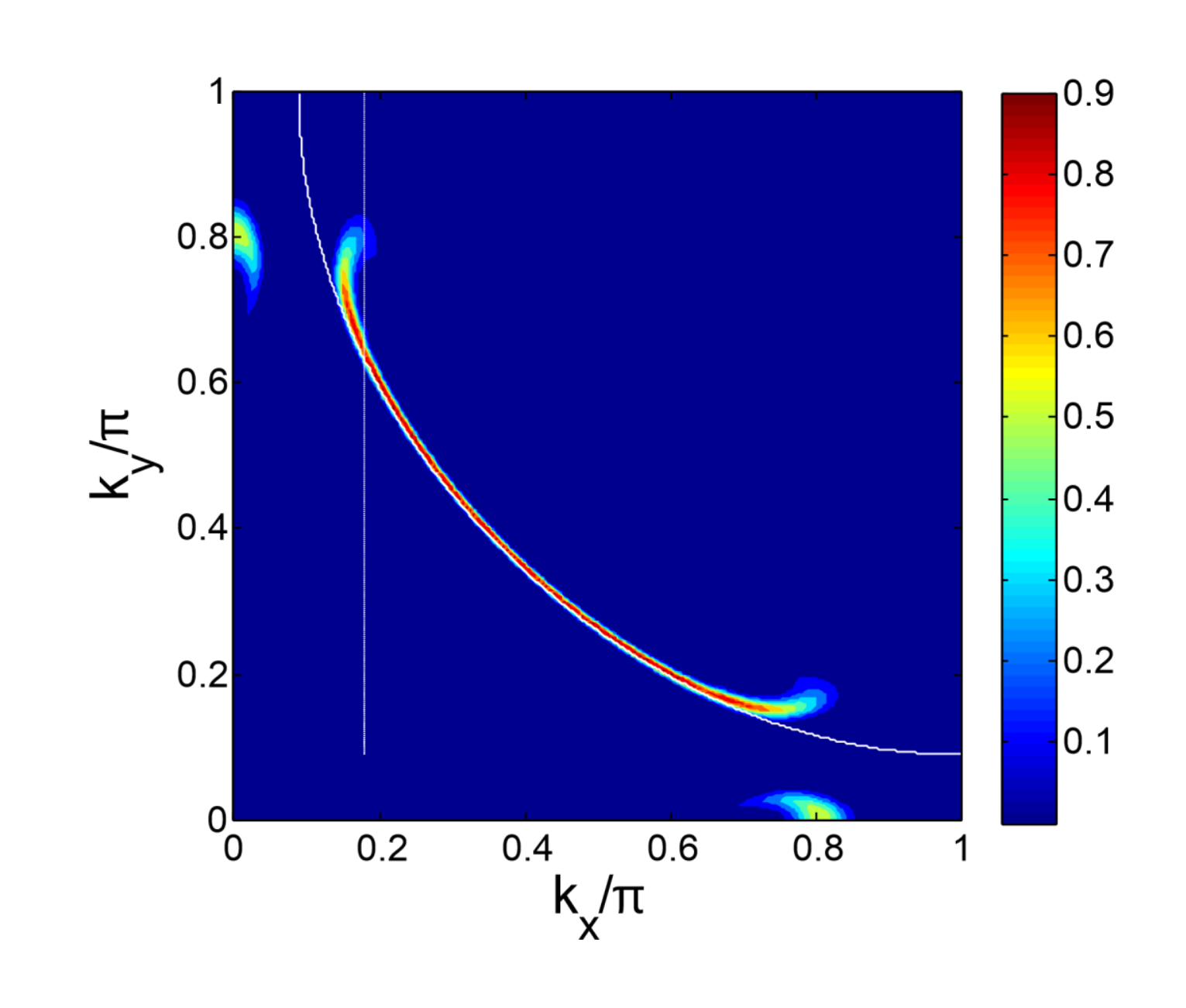}
\caption{ The spectral weight of the bands (gaussian broadened with a 10 meV width) at the Fermi level. Parameters are the same as in Fig. 3, which correspond to a doping density p=0.207. The white line shows the bare Fermi surface. The vertical line marks $k_x$=Q. Note that the tip of the Fermi arc is close to this line, hence the wavevector connecting the Fermi arcs has length near 2Q, which is the CDW wavevector in our theory.}
\end{figure}

In Fig. 5 we show the tunneling density of states

\begin{figure}[h]
\includegraphics[width=3.25in]{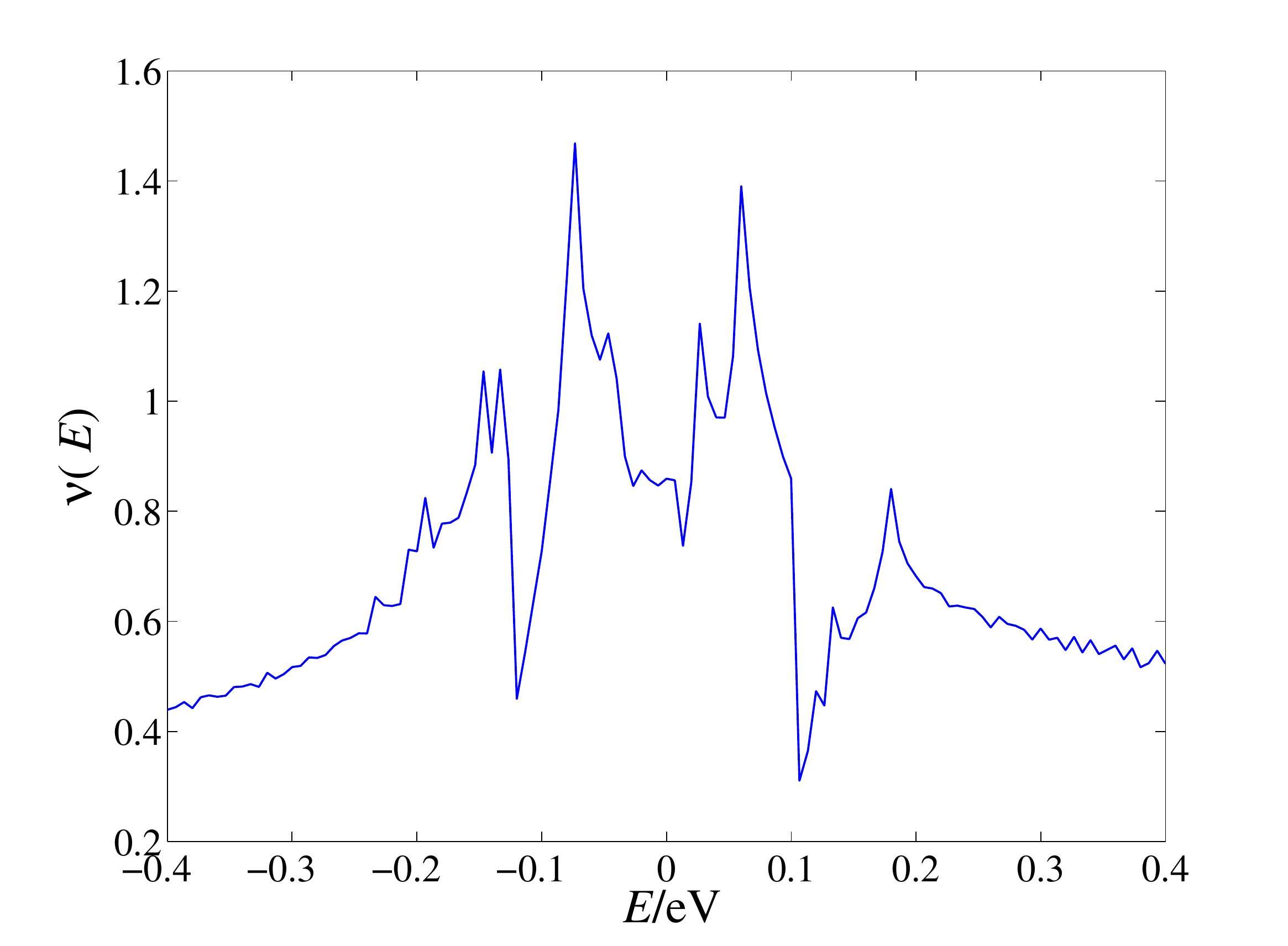}
\caption{Plot of the tunneling density of state $\nu(E)$ for the same parameters as Fig.3.}
\end{figure}

\begin{equation}
\nu (E) = \sum_{\bm k,\alpha} | \nu_\alpha |^2 \delta (E-E_\alpha) .
\end{equation}
It shows a pseudogap with some asymmetry. After lifetime broadening, it resembles STM tunneling data.

The density operator 
$
\rho_{2\bm Q} = \sum_{\sigma}c^\dagger_{\bm k - 2\bm Q,\sigma} c_{\bm k,\sigma}
$
has a nonzero average in mean field theory, i.e. $< \rho_{2\bm Q} >$ is an induced order parameter.  Its physical interpretation in the case of spin liquid is a modulation of the singlet valence bond strength, i.e., an incommensurate valence bond solid.\cite{14}  For the superconductor this gives rise to a CDW.   Our theory predicts the CDW vector to be twice the PDW vector, which we set to be the vector spanning the Fermi surface at the anti-nodes.  We have focused on $(0,\pi)$ where the spanning vector is $\bm Q_1 = (Q,0)$.  Similar consideration near $(\pi,0)$ gives $\bm Q_2 = (0,Q)$.  Since the two anti-nodal regions are far apart and relatively independent, we expect the pairing at the two nodes to be weakly coupled and co-exist.  Thus our picture favors the bi-directional CDW (checkerboard) rather than uni-directional stripes, in agreement with the pattern seen by STM on Bi-2201.\cite{6}  In~Fig.6 0we compare the predicted CDW ordering vector $2Q$ with a collection of measurements on Bi2201 and the agreement is satisfactory. Note that in our theory 2Q is also correlated with the wavevector connecting the tip of the arcs, in agreement with the empirical observation by Comin et al. \cite{5} It will be good to make a similar comparison using the data on Bi2212, \cite{6b} if accurate data on the antinodal spanning vectors are available.
\begin{figure}
\includegraphics[width=4.25in]{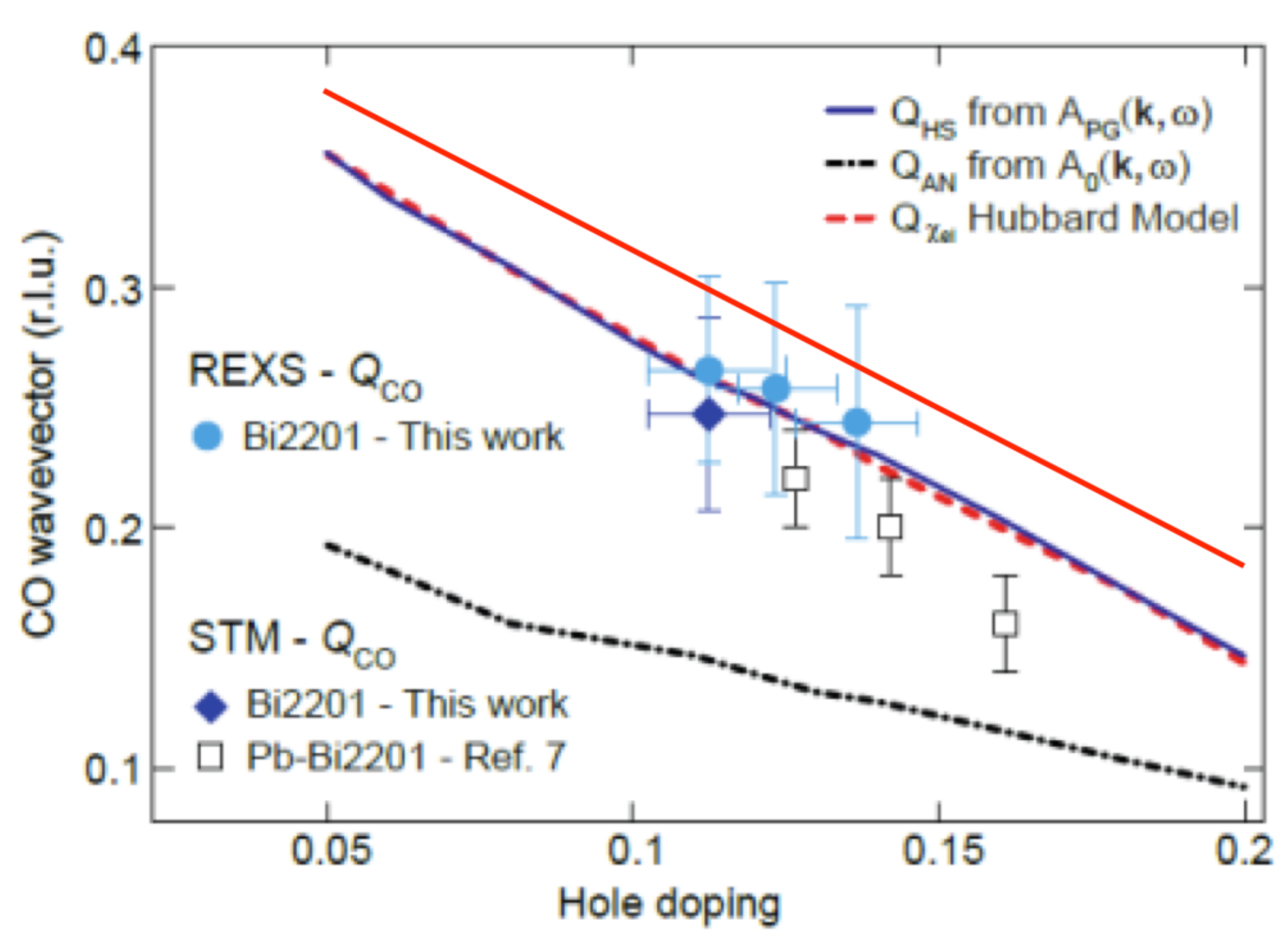}
\caption{A collection of data for the CDW wavevector $\delta$ (in units of $2\pi$) vs hole doping concentration on the Bi2201 system, taken from ref.\cite{5}.  The dashed dotted line marked $Q_{\rm AN}$ is the Fermi surface spanning vector at the anti-node which corresponds to $Q$ in our notation.  The solid red line is approximately $2 Q_{\rm AN}$ which is the prediction of the Amperean pairing model.}
\end{figure}

In materials with intrinsic anisotropy due to chains such as YBCO, it is found that the CDW scattering integrated intensity is stronger when $\bm Q$ is parallel to the chains by  a factor which is at most 2.5 in intensity, i.e. 1.6 in $\rho_{2\bm Q}$.\cite{26,27} This anisotropy is presumably responsible for the nematic behavior in transport measurements such as Nernst effect. \cite{8} We believe that CDW with both $\bm Q_1$ and $\bm Q_2$ co-exist in a given region, but with different strength.  As discussed below, this co-existence is needed to understand the quantum oscillation data.

By leaving segments of the Fermi surface ungapped, the Amperean paired state has a finite normal fluid density even at zero temperature and hence a smaller superfluid density $\rho_s$ compared with the uniform state.  We assume phase fluctuations and the nucleation of vortex anti-vortex pairs suppresses phase coherence, so that pairing is only short range ordered until the conventional $d$-wave order takes over and gaps out the Fermi arc.  The competition for the remaining Fermi surface leads to the reduction of Amperean pairing below $\rm T_c$, as clearly seen in X-ray experiments.  
A schematic phase diagram is shown in Fig.7, where a short range ordered PDW state is the dominant feature.  
As a competing phase, it is natural for this state to form a large vortex core when the $d$-wave superconductor is subject to a magnetic field.  This explains the observation of checkerboard patterns near the vortex core by STM some time ago.\cite{28} At a relatively low field these cores overlap. As shown in Fig.7(b), the small critical field \cite{11} marks the transition to the short range ordered Amperean pairing state. 
The phase diagram also shows a long range ordered CDW state forming out of the short range ordered PDW, as explained in the next section.  In real materials the CDW is short range ordered due to disorder pinning.
We follow Harrison and Sebastian \cite{29} and use the $2\bm Q_1$ and $2\bm Q_2$ vectors to connect pieces of the Fermi arcs, giving rise to a small pocket which may be the origin of the quantum oscillations with electron-like carriers.  The recent observation of quantum oscillation and CDW in the Hg1201 compound found a pocket area which is larger than that of YBCO while the CDW wave vector is shorter. \cite{30} The trend supports this scenario.

\begin{figure}
\includegraphics[width=3.25in]{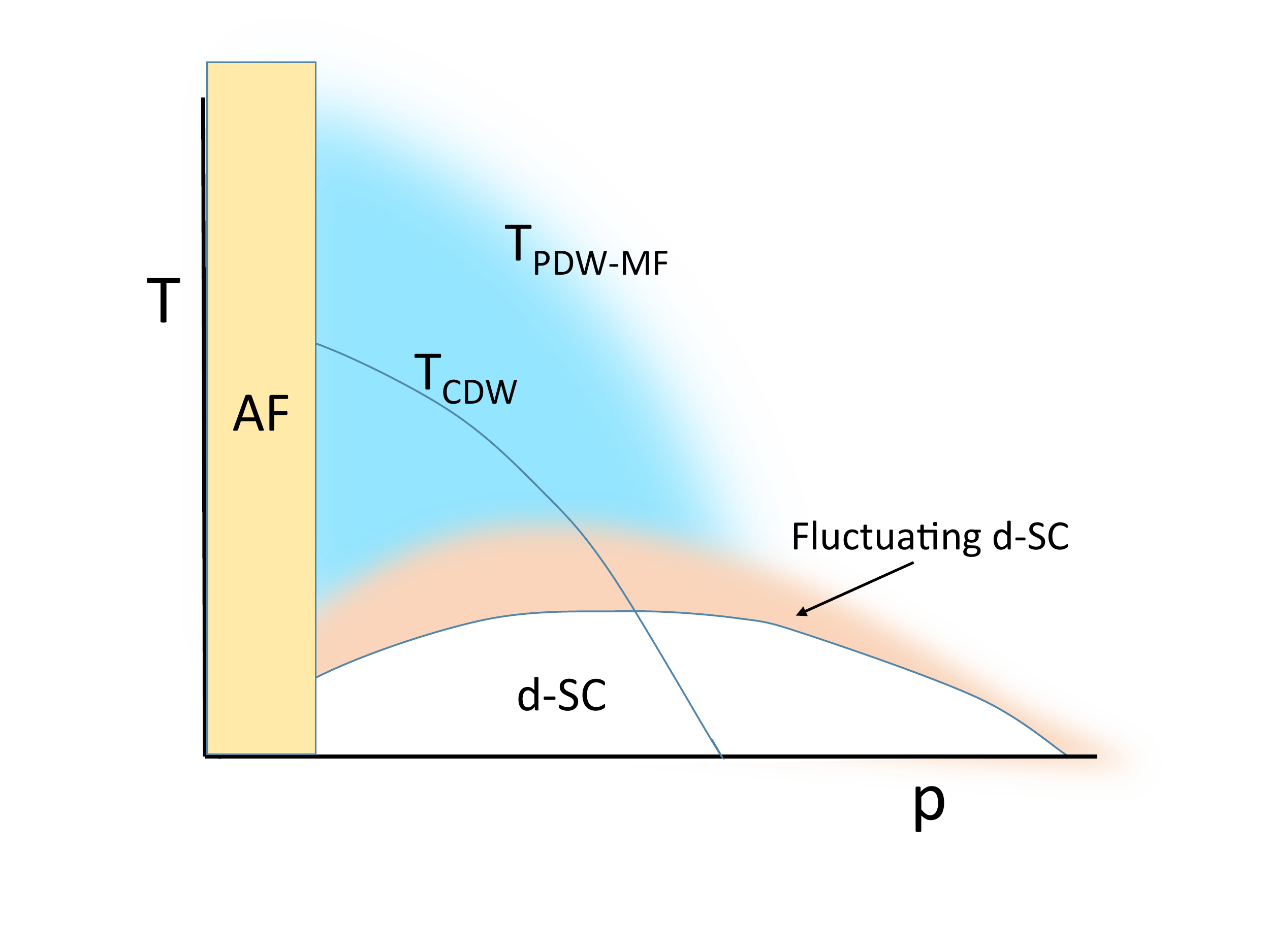}\includegraphics[width=3.25in]{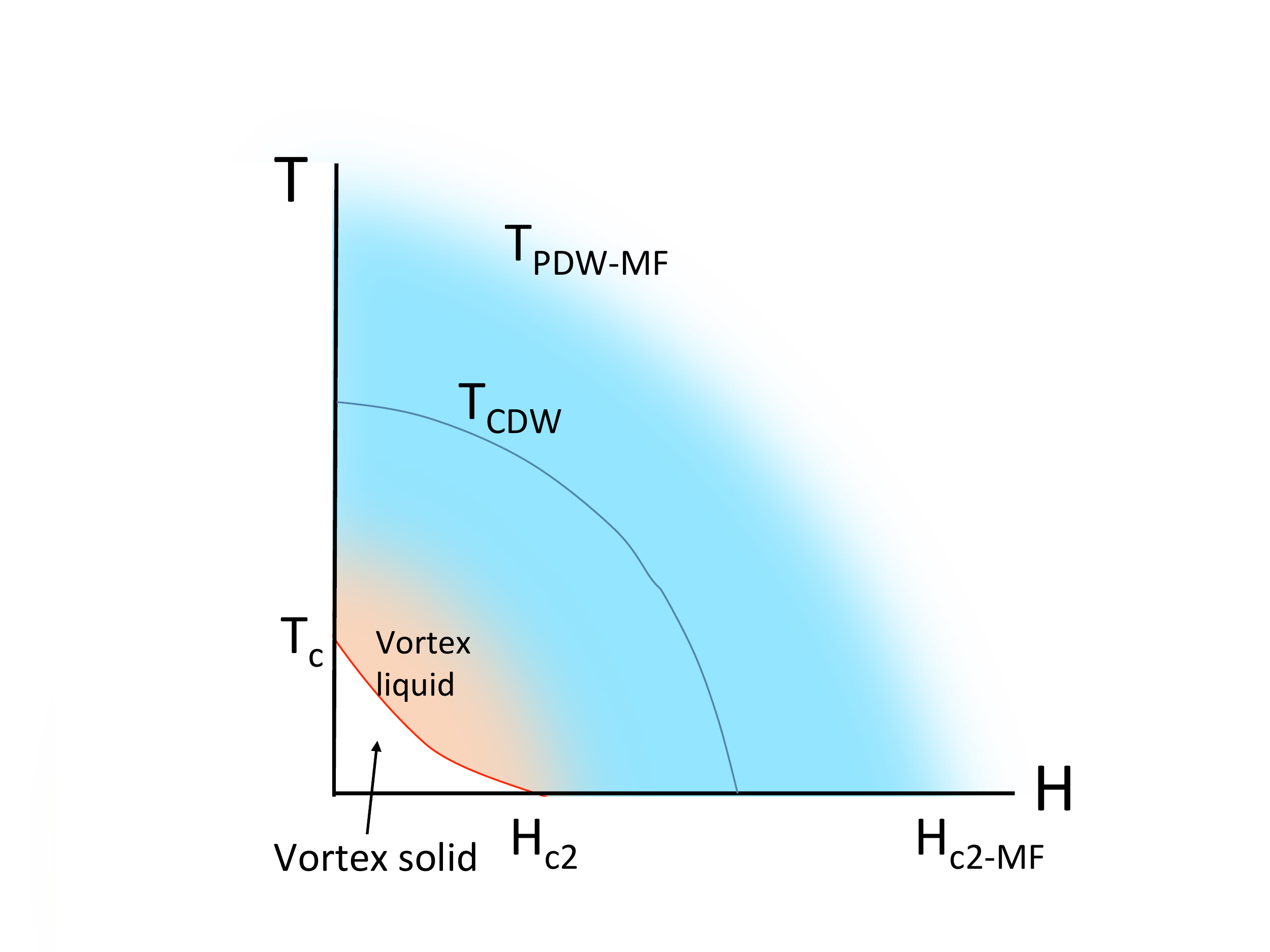}
\caption{Proposed phase diagram.  (a) $T$ vs doping concentration $p$.  $T_{{\rm PDW-MF}}$ is a cross-over temperature where the pairing amplitude is established.  The phase diagram is dominated by the shaded blue region representing PDW superconductivity which is short range ordered due to strong phase fluctuation.  It is responsible for the pseudo gap.  CDW order appears below $T_{{\rm CDW}}$.  In real material disorder pinning destroys true long range order.  $d$-wave superconductivity appears below ${\rm T_c}$ accompanied by fluctuating $d$-wave superconductivity which is tied to $T_c$.  
The PDW, either fluctuating or with long range order, co-exists with $d$-wave pairing up to some doping.
The precise location and nature of the antiferromagnetic (AF) boundary depends on disorder. (b) $T$ vs magnetic field $H$.  The red line is the vortex melting temperature which separates the vortex solid from the vortex liquid.  At zero temperature for $H > H_{c2}$, vortex cores overlap, resulting in a short range order PDW state which is metallic and shows quantum oscillations.  The CDW order is destroyed beyond a certain magnet field. }
\end{figure}
Finally we discuss a possible explanation of the appearance of Kerr rotation below T$^\ast$.  An important recent development is the realization that the Kerr effect is due to a gyrotropic effect which requires the breaking of reflection symmetry in all directions, as opposed to the breaking of time reversal symmetry as originally assumed.\cite{31}  Ref. \cite{31} offers several examples when $Q$ is commensurate.  Here we present a phenomenological scenario applicable to the incommensurate case.

Let us define the phases of the various order parameters as $\Delta_{\pm \bm Q_\alpha} = | \Delta_{\pm \bm Q_\alpha}|  \exp \left(  i\theta_{\pm \bm Q_\alpha} \right) $, $\rho_{2\bm Q_\alpha} = | \rho_{2\bm Q_\alpha} | \exp (i\phi_\alpha), \alpha = 1,2$.  If $| \Delta_{\bm Q_\alpha} | = |\Delta_{-\bm Q_\alpha} |$ the pairing is modulated in space as
\begin{equation}
\Delta (\bm r) = \sum_\alpha | \Delta_{\bm Q_\alpha} | e^{i\tilde{\theta}_\alpha} \cos
\left(
\tfrac{1}{2}(\theta_{\bm Q_\alpha} - \theta_{ -\bm Q_\alpha}) + \bm Q_\alpha \cdot \bm r
\right)
\end{equation}
where $\tilde\theta_\alpha = \tfrac{1}{2} (\theta_{\bm Q_\alpha} + \theta_{ -\bm Q_\alpha})$ plays the role of the overall pair phase while $\theta^\prime_\alpha = \tfrac{1}{2} (\theta_{\bm Q_\alpha} - \theta_{-\bm Q_\alpha} )$ determines the location of the crests of the PDW.   The term in the free energy
\begin{eqnarray}
F &=& a \left\{
\rho_{2\bm Q_1} \Delta^\ast_{\bm Q_1} \Delta_{-\bm Q_1} + \rho_{2 \bm Q_2} \Delta^\ast_{\bm Q_2} \Delta_{\bm Q_2} + c.c. \right\} \nonumber \\
\\
&+& a^\prime \left\{
\rho_{\bm Q_1 + \bm Q_2} \Delta^\ast_{\bm Q_1}\Delta_{\bm Q_2} + c.c. 
\right\} \nonumber
\end{eqnarray}
is the phenomenological basis for generating the induced order $\rho_{2\bm Q_1},\rho_{2\bm Q_2}$ and $\rho_{\bm Q_1 + \bm Q_2}$.\cite{14},\cite{31a},\cite{17} The $\bm Q_1 + \bm Q_2$ order has so far not been observed.  Eq.(8) generates locking between the phases of the form $\cos (\theta_{\bm Q_\alpha} - \theta_{-\bm Q_\alpha} - \phi_\alpha)$. $\tilde\theta_\alpha$ and $\theta^\prime_\alpha$ appears to be independent, but there are subtleties associated with vortex formation.\cite{17},\cite{31a} Nevertheless it is possible  that phase fluctuations produced by conventional $hc/2e$ vortices can destroy the $\tilde\theta_\alpha$ correlation while leaving $\theta^\prime_\alpha$ intact. Thus it is possible that the induced CDW order has long range 
order while the pairing is short range.\cite{23a} In reality, disorder pinning limits the spatial order.  For incommensurate $Q$, $\theta^\prime_\alpha$ can be set to zero by choosing the origin of $\bm r$ and the locking term becomes  simply  $\cos\phi_\alpha$.   Physically $\phi_\alpha$ describes the relative location of the crests of the PDW and CDW.  When it is locked to 0 or $\pi$, reflection symmetry is preserved in the plane.  A deviation of $\phi_\alpha$ from 0 or $\pi$ can come from higher order terms in the free energy of the form $b (\rho_{2\bm Q_\alpha} \Delta_{\bm Q_\alpha} \Delta^\ast_{-\bm Q_\alpha})^2 + c.c.$ etc., which gives a locking term $b \cos 2\phi_\alpha$.  If $a$ and $b$ are opposite in sign and $|b| > |a|/4$ the free energy has a minimum at $\phi_\alpha = \pm \phi_0$ thereby breaking the reflection symmetry in the plane.  To break reflection symmetry along $z$ note that the stacking in the $z$ direction can take on two distinct forms: $(\phi_1,\phi_2) = [ (\phi_0,\phi_0), (\phi_0, -\phi_0), (-\phi_0,-\phi_0), (-\phi_0,\phi_0)  ]$ or
$(\phi_0,\phi_0), (-\phi_0, \phi_0), (-\phi_0,-\phi_0), (\phi_0,-\phi_0)]$, thereby introducing a sense of chirality.  While we do not have a microscopic basis for the appearance of a finite $\phi_0$, this argument at least offers the possibility that Kerr rotation onset is associated with Amperean pairing and checkerboard CDW order.

Is there a way to directly detect the PDW? The pairing response function $\chi (q,\omega)$ can be measured \cite{32} by constructing a tunnel junction between an optimally doped cuprate superconductor and a material with a low {\rm T$_c$}, such as Bi-2201 and operate at a temperature between the two {\rm T$_c$}'s. The pair momentum is supplied by a parallel magnetic field. The tunneling current vs magnetic field and voltage is predicted to be proportional to $Im\chi (q,\omega)$ where $\omega =2eV/\hbar$ and $q=(2eB/\hbar c)(\lambda + d/2) $ where $\lambda $ is the penetration depth of the superconductor (assumed to be thick) and d the barrier thickness.  The fluctuating PDW will give rise to a peak in the tunneling current at $B$ corresponding to $q = \pm Q$. The same experiment can be performed at low temperature using d wave or conventional s-wave superconductor. If the PDW co-exists with d wave pairing, a peak in the current is predicted to emerge at a field far greater than that expected for the Fraunhofer pattern coming from the d-wave order. The shape of this peak will contain information about the degree of the PDW order. In fact, the PDW may even develop long range order (limited by disorder). The reason is that a term  $(\Delta^\ast_0)^2\Delta_{\bm Q_\alpha} \Delta_{-\bm Q_\alpha}$ is allowed, where $\Delta_0$ is the d wave pairing order parameter. This term pins the phase $\theta_{\bm Q_\alpha} + \theta_{ -\bm Q_\alpha}$ and since $\theta_{\bm Q_\alpha} - \theta_{ -\bm Q_\alpha}$ is assumed to be already locked, the individual phases $\theta_{\bm Q_\alpha}$ will be pinned up to $\pi$. Thus  $\Delta_{\bm Q_\alpha}$ may have long range order, in the phase where domain walls with $\pi$ phase shifts are not important.   In this case, the term in the Landau Free energy $\rho_{\bm Q_\alpha} \Delta_{\bm Q_\alpha} \Delta^\ast_0$ implies that a CDW with wave-vectors $\bm Q_\alpha$ is predicted to appear in the superconducting state.

It has come to our attention that STM measurements at 6K on optimally doped YBCO has found charge ordering with two sets of ordering vectors at $\delta$ = 0.28 $\pm 0.03$ and 0.14 $\pm 0.01$ ( in units of $2\pi$) along the x and y axes. \cite{46} We would like to re-interpret these CDW's to correspond to our $\rho_{2\bm Q_\alpha}$ and $\rho_{\bm Q_\alpha}$ respectively. The recent discovery of CDW order at $\delta$ = 0.28 in optimally doped BISCO lends support to this interpretation. \cite{47} If confirmed by direct X-ray measurement, the appearance of $\rho_{\bm Q_\alpha}$ in the superconducting state serves as a strong confirmation of our theory. 

In conclusion, recent experimental advances have put severe constraints on the nature of the pseudo gap state.  The assumption of Amperean pairing gives a consistent account of all the unusual phenomena.  It remains to be fully understood why long range order is not achieved below $T^\ast$ and above a small ``$H_{c2}$.''  Phase fluctuations are presumably at play but a detailed thermal and quantum description of the short range ordered state will be highly desirable.  If it is possible to increase the interlayer Josephson tunneling by building artificial MBE structures, we expect phase fluctuations will be suppressed and it will be extremely interesting to see if a long range ordered Amperean superconducting state at a relatively high temperature between $T_c$ and $T^\ast$ can be stabilized. 

I thank N. Phuan Ong for emphasizing to me the pairing nature of the pseudo gap phase and for sharing his insight that the high field state is some form of PDW.  I also thank T. Senthil for many discussions on the high $\rm T_c$ topic.  I acknowledge support by NSF grant DMR--1104498.

\clearpage
\appendix
\section{Incompatibility of the CDW model with the ARPES data.}
With the discovery of CDW order, it seems natural to associate the energy gap induced by CDW with the pseudogap.\cite{S1,S2,S3,S4}  Here we show that a mean field picture of the CDW fails to explain the ARPES data of He {\em et al}.\cite{7}  Figure 8 shows several scans of the spectrum along $k_x$ starting from the anti-node as indicated in Fig.1.  By a judicial choice of the CDW ordering vector $\delta$ and the gap size, the scan at the anti-node can account for the ARPES data reproduced in Fig.3(b).  However, the agreement breaks down away from the anti-node.  Figure 8(b) and (c) show the spectrum for $\delta k_y$ near the appearance of the Fermi arc, and in the middle of the Fermi arc where the band crosses the Fermi level.  It is clear that the Fermi arc is formed by a state moving {\em down} towards the Fermi level, leaving a large gap just below.  This is in strong contradiction with the data shown in Fig.3(d), which shows that the gap is closed by a state moving up in energy to meet the Fermi level.  In particular, the CDW model predicts a large gap below the Fermi level at the end of the Fermi arc, which has never been seen experimentally. 
In fact, it has been emphasized that a gap exists {\em above} the Fermi level near the end of the Fermi arc.\cite{25}
\begin{figure}[t]
\includegraphics[width=3.25in]{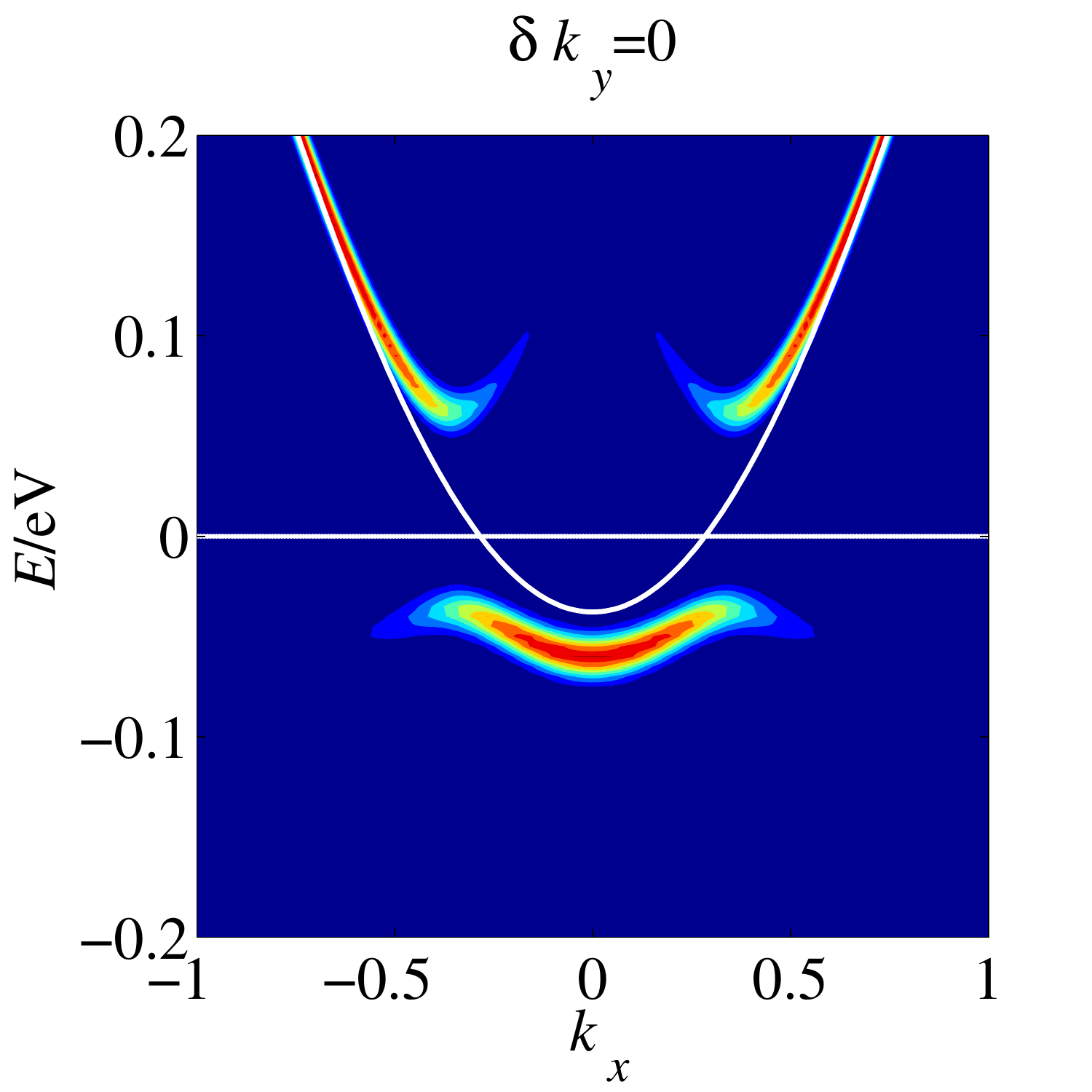}\includegraphics[width=3.25in]{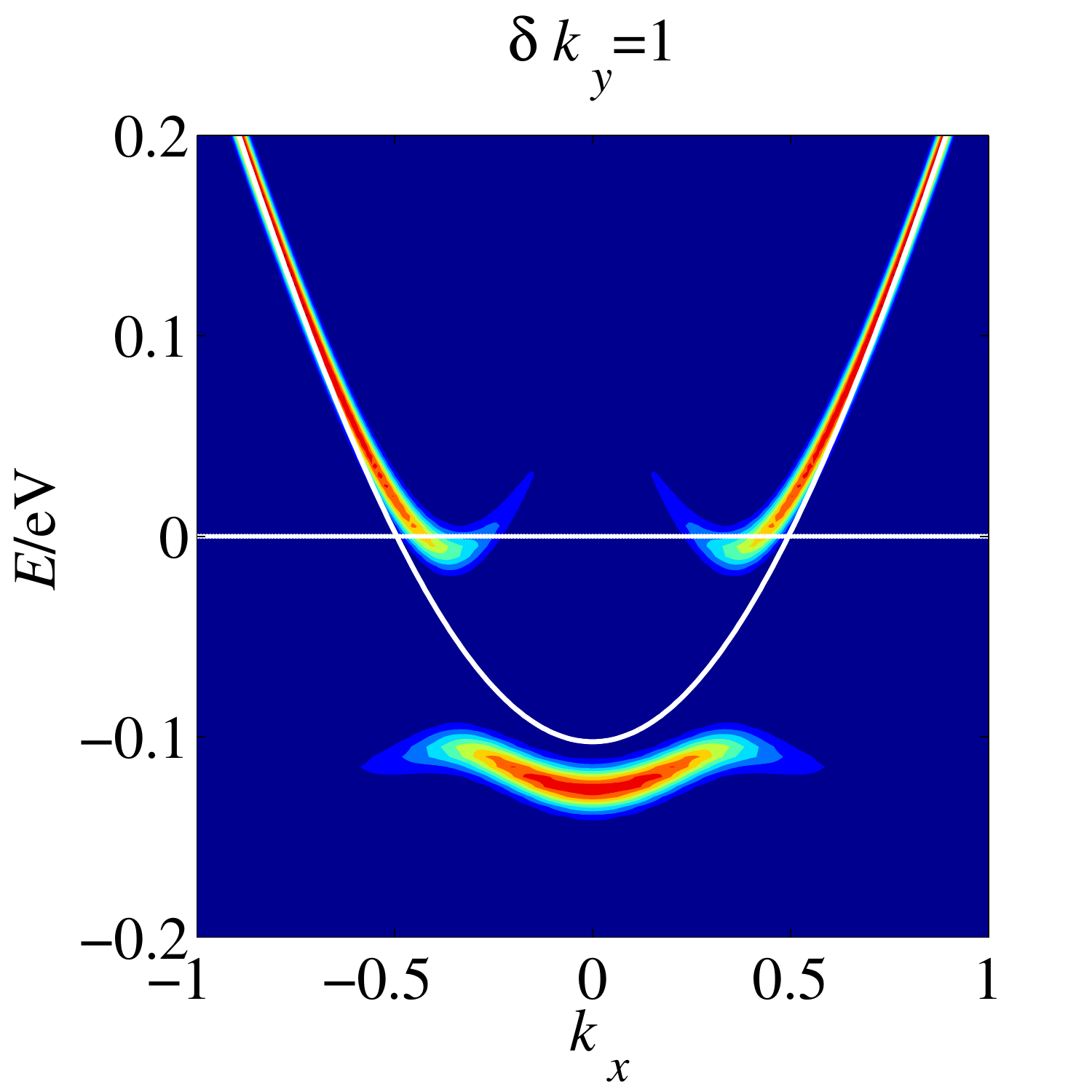}
\includegraphics[width=3.25in]{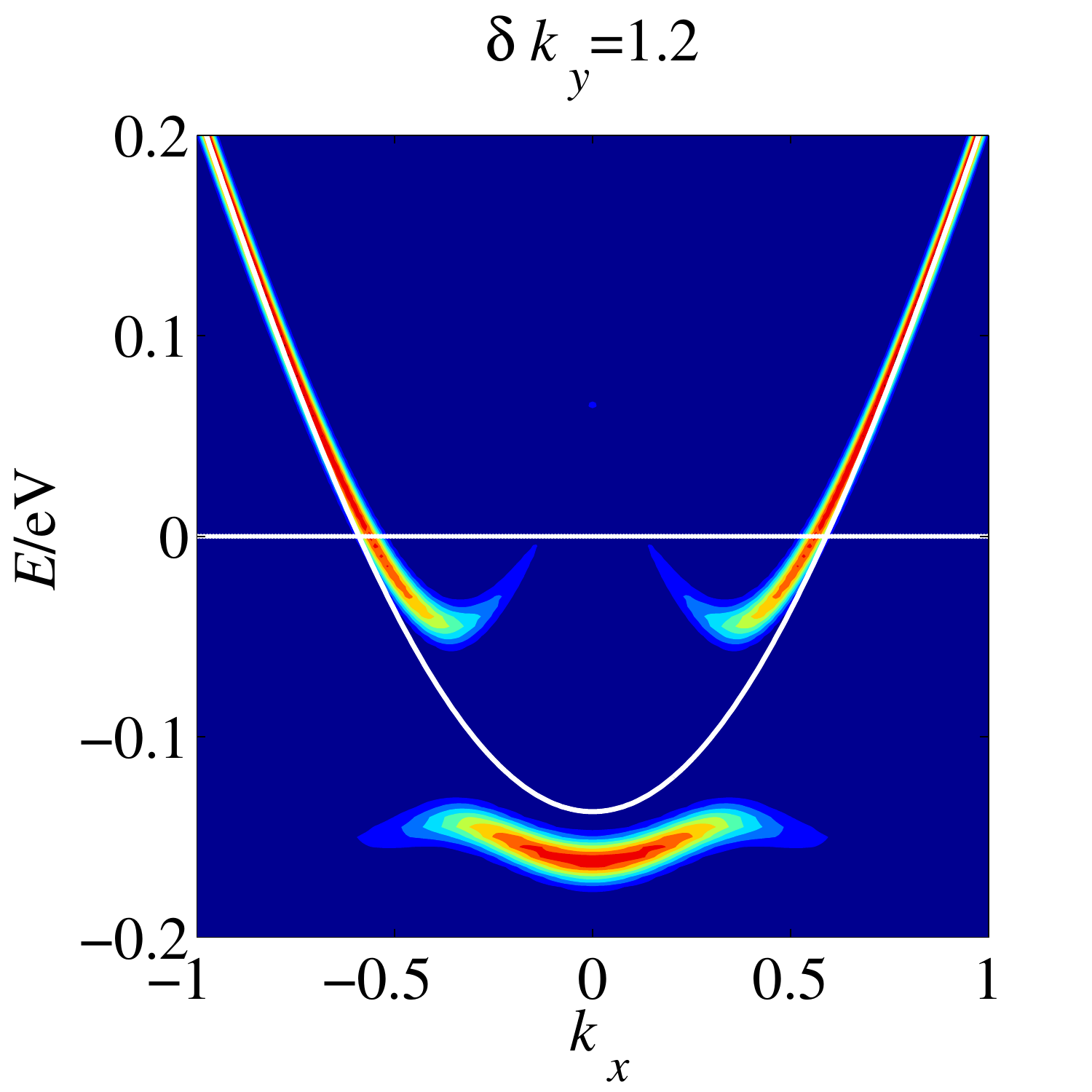}
\caption{Plot of the energy spectrum vs $k_x$ for (a) $k_y = \pi$, $\delta k_y = 0$, (b) $\delta k_y = 1.0$ and (c) $\delta k_y = 1.2$.  The CDW wave vector is assumed to be $1.2 Q$ and the gap is set to be 0.05~eV.  Fig.(c) corresponds to a cut through the Fermi arc while Fig.(b) is a cut near the end of the arc.  Note that the state at the Fermi level arises by a state moving from above, leaving a gap below the Fermi level, in strong disagreement with the data shown in Fig.3(b)-(e).}
\end{figure}

\begin{figure}[h]
\includegraphics[width=3.25in]{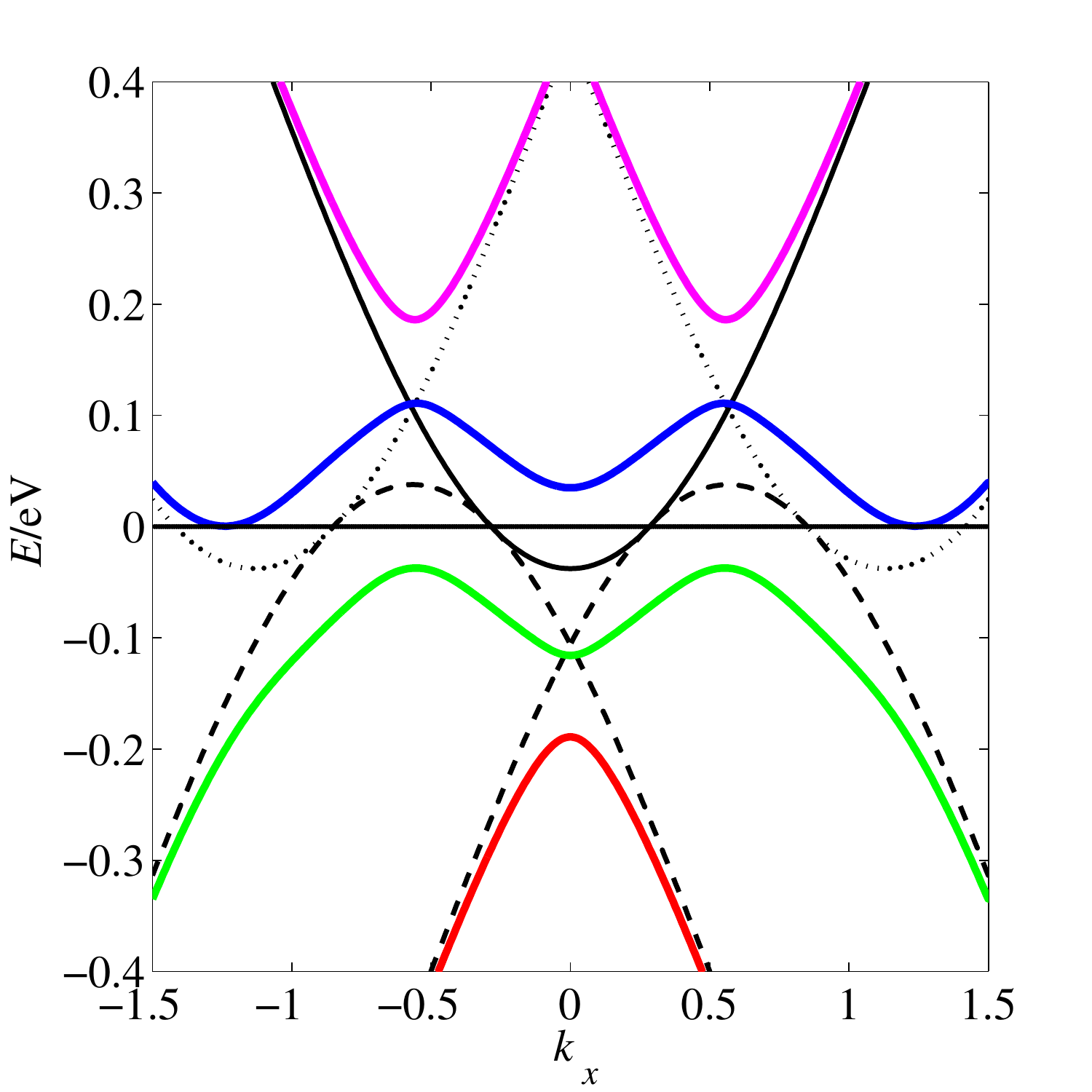}\includegraphics[width=3.25in]{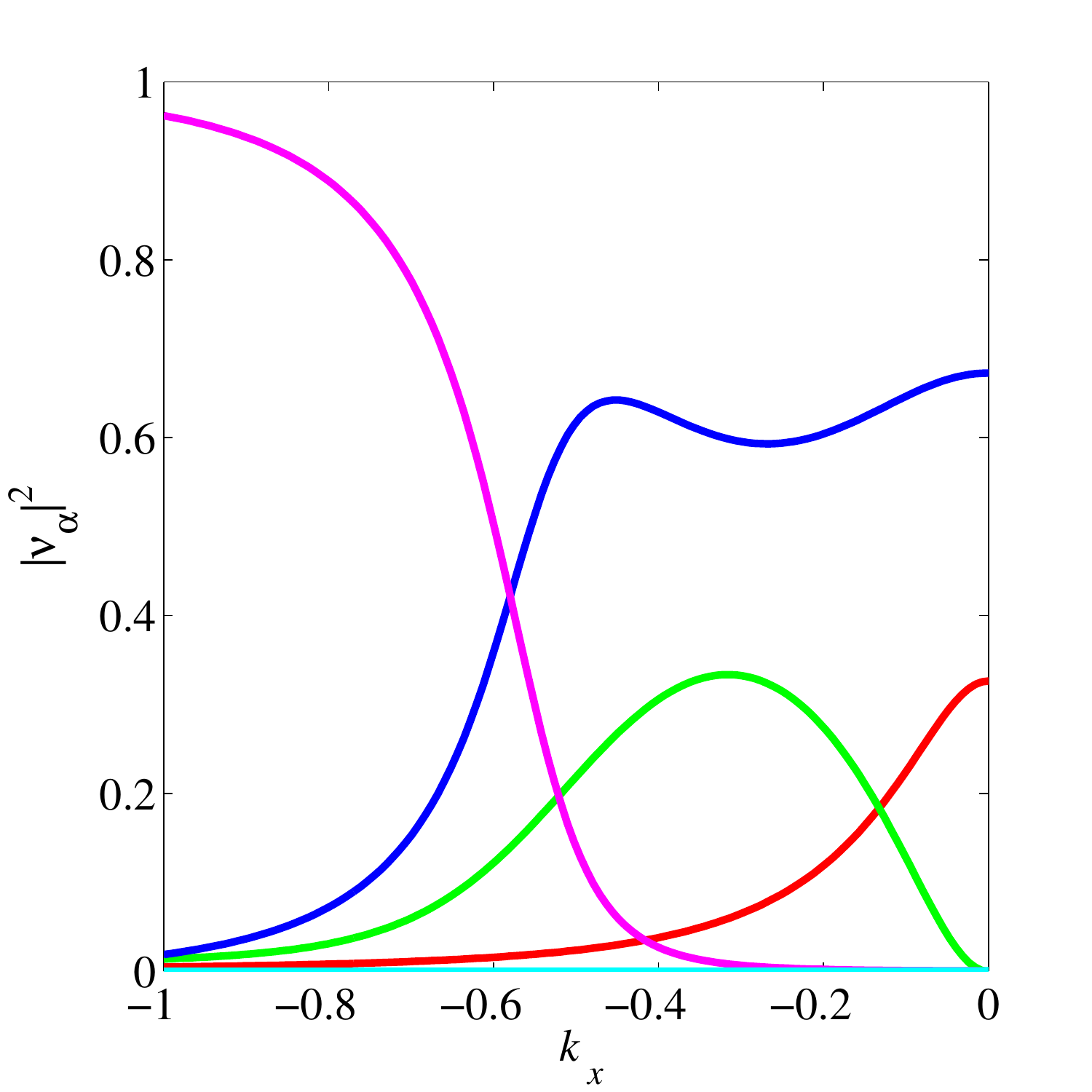}
\caption
{(a) Plot of the energy bands of the five band model defined by Eq.(5) for $k_y$=$\pi$. The dashed lines shows $-(\varepsilon_{-\bf k \pm {\bf Q}}-\mu)$. The dotted line shows  $(\varepsilon_{\bf k \pm 2{\bf Q}}-\mu)$. (b) The spectral weight of the bands. The parameters used are the same as in Fig.3, with the band parameters taken from ref \cite{7}.}
\end{figure}

\begin{figure}[h]
\includegraphics[width=3.25in]{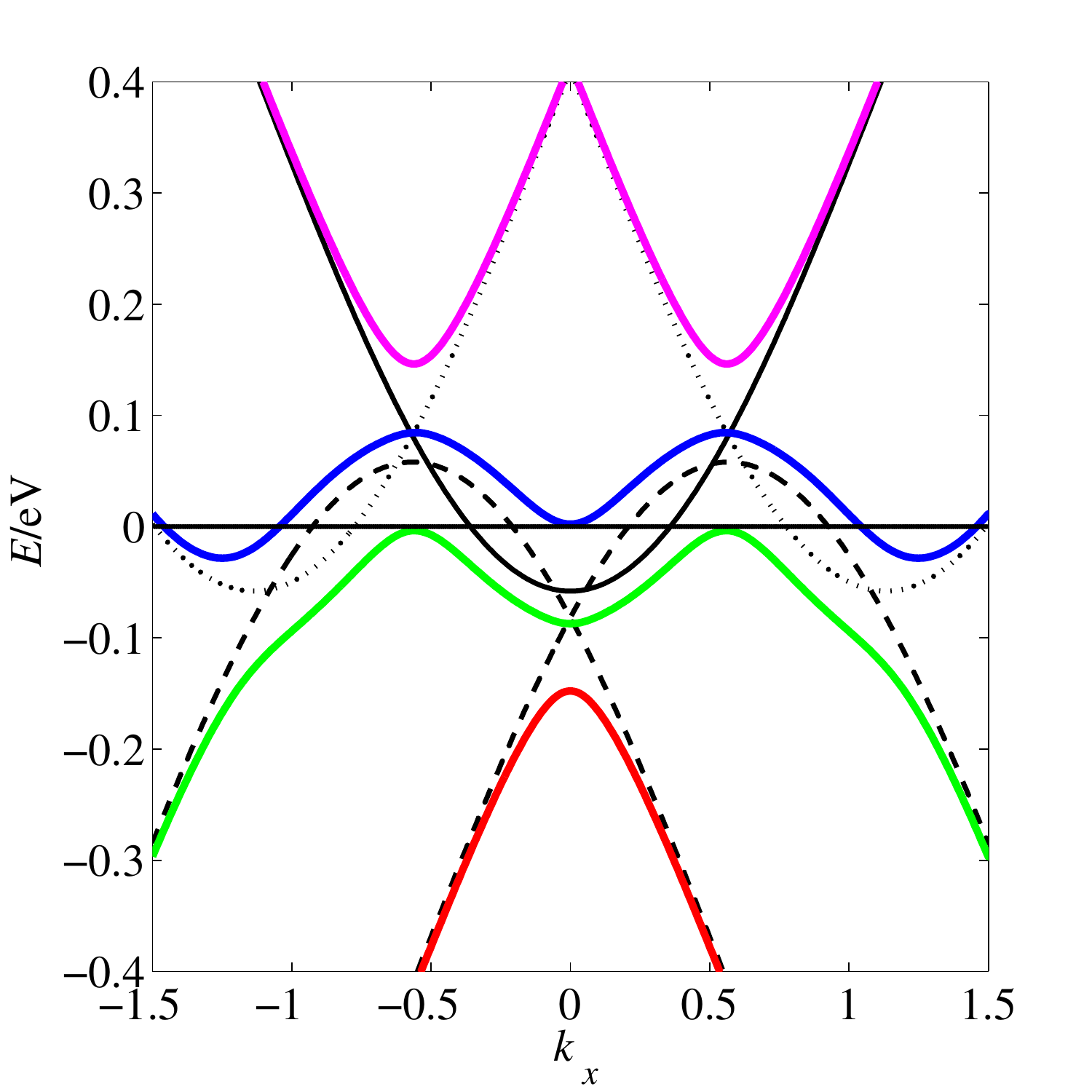}\includegraphics[width=3.25in]{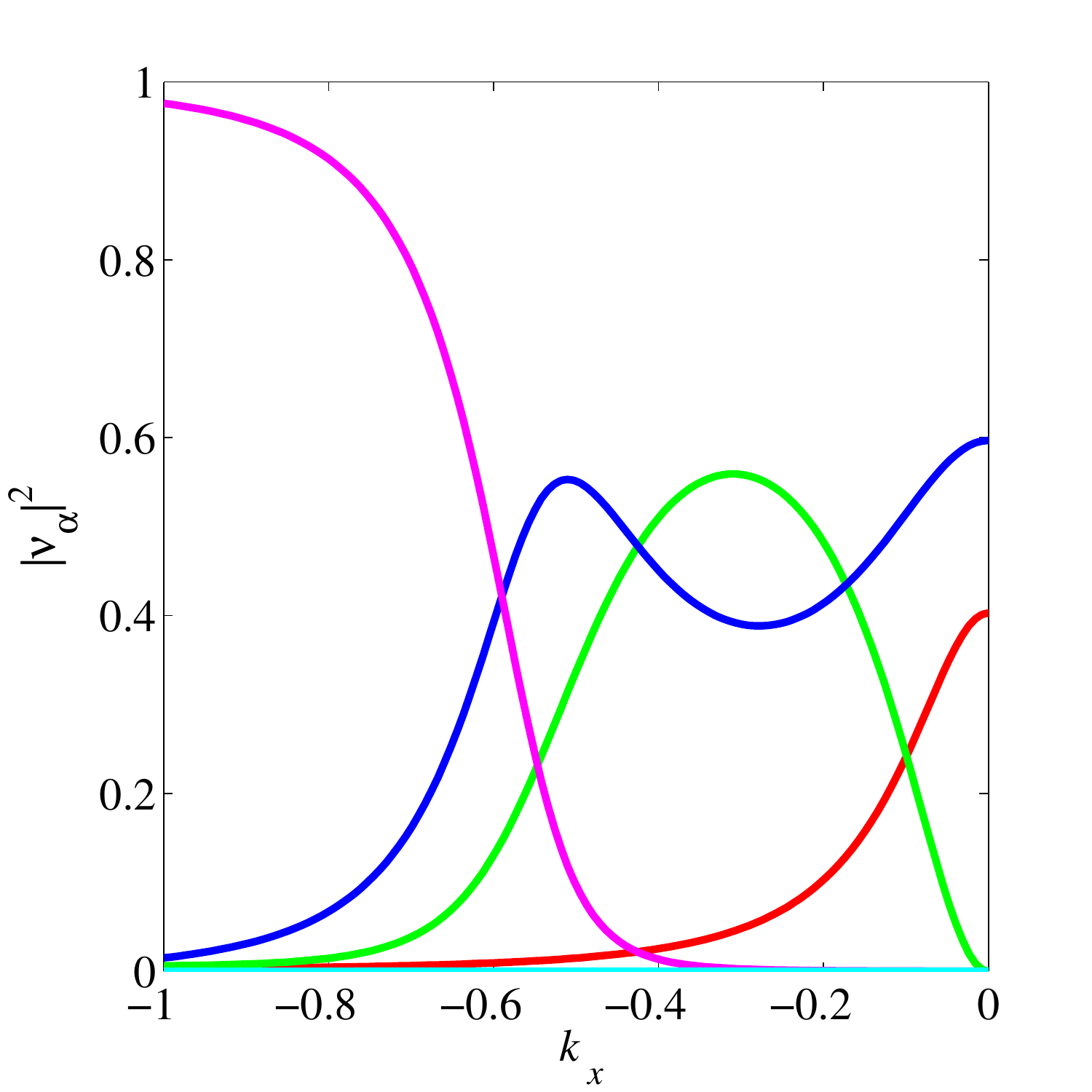}
\caption{Same as Fig.9 except that $\delta k_y$=0.6 }
\end{figure}

\begin{figure}[h]
\includegraphics[width=3.25in]{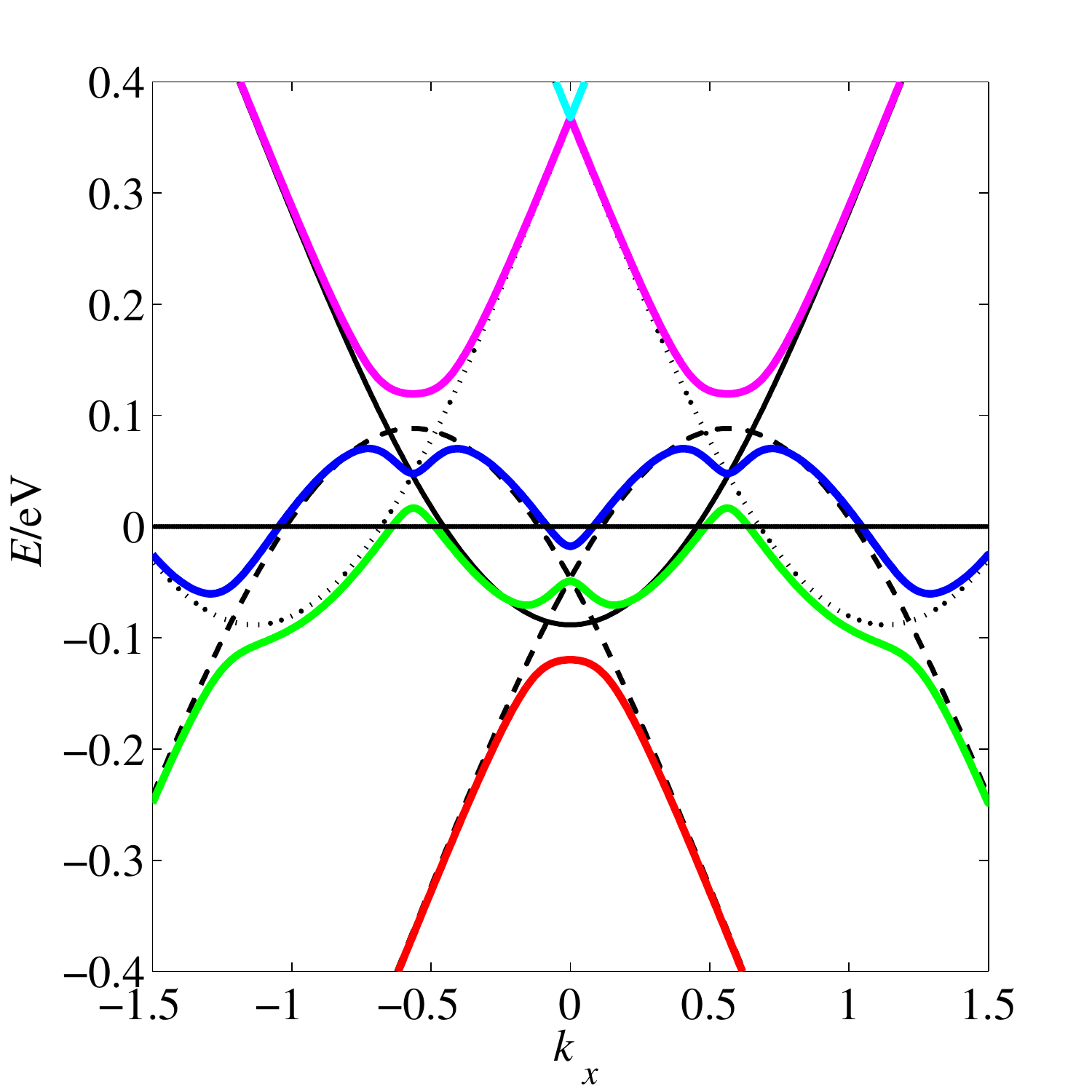}\includegraphics[width=3.25in]{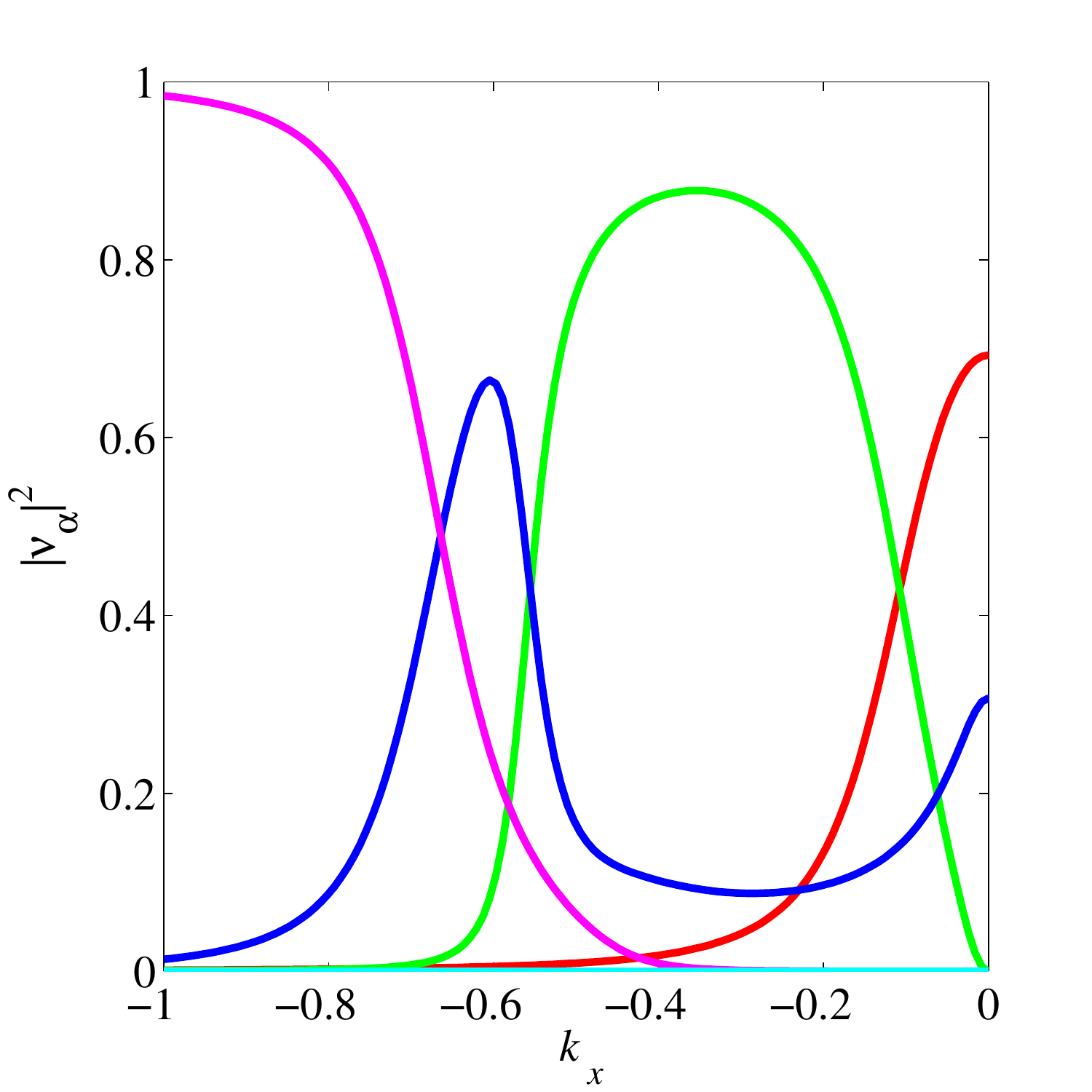}
\caption{Same as Fig.9 except that $\delta k_y$=0.9 }
\end{figure}

We note that ref. 51 assumes an interference between the CDWs in the $x$ and $y$ directions, so that the gap vanishes at the ``hot spot.''  However, the gap reopens away from the ``hot spot'' and 
should be visible below the Fermi level in the Fermi arc region.
 Alternatively in the bond density wave picture, the gap is ${\bm k}$ dependent and vanishes at the node.\cite{S1,S2,S3}  However, the difficulty described here remains as long as the gap is nonzero at the end of the Fermi arc.

\section{Details of the energy spectra}
We show in Fig.9 the dispersion and the spectral weight of the solution of the 5 band model  defined by Eq.(5) for $k_y$=$\pi$.  Note the common tangents of the original bare band with the hole bands (solid an dashed lines) which give rise to large gaps at the Fermi level. At $k_x$=0, the bare hole bands cross at -100meV. One linear combination is decoupled from the electron band and the dispersion of the geen band goes through this point, albeit with zero spectral weight because it is decoupled from the electron. This feature can also be seen in the 3 band spectrum shown in Fig 2.  Figure 10--12 shows how the band structure evolves as $k_y$ deviates from $\pi$. The top of the green band moves towards the Fermi level and crosses it near $\delta k_y$ = 0.6. Beyond that it forms what looks like a pocket [see Fig.11(a)] but  the back side of the pocket is 
mainly a hole-like quasiparticle, so that its spectral weight for removing an electron is 
so small that it is not visible in Fig. 4. Instead what is seen is a bending of the "Fermi surface" away from the bare Fermi surface. (see Fig. 4). We also note that due to the repulsion by the electron bands shifted by $\pm 2Q$ (dotted lines) which crosses the original bare band at  $k_x=\pm Q$ ,the top of the green band is located near  $k_x=\pm Q$. As a result, the tip of the Fermi arc as defined by the touching of the green band to the Fermi level lies close to $k_x=\pm Q$. Consequently the spanning vector of the tips of the arcs is 2Q, which is also the CDW wavevector $\delta$ in our theory.
 \begin{figure}[h]
\includegraphics[width=3.25in]{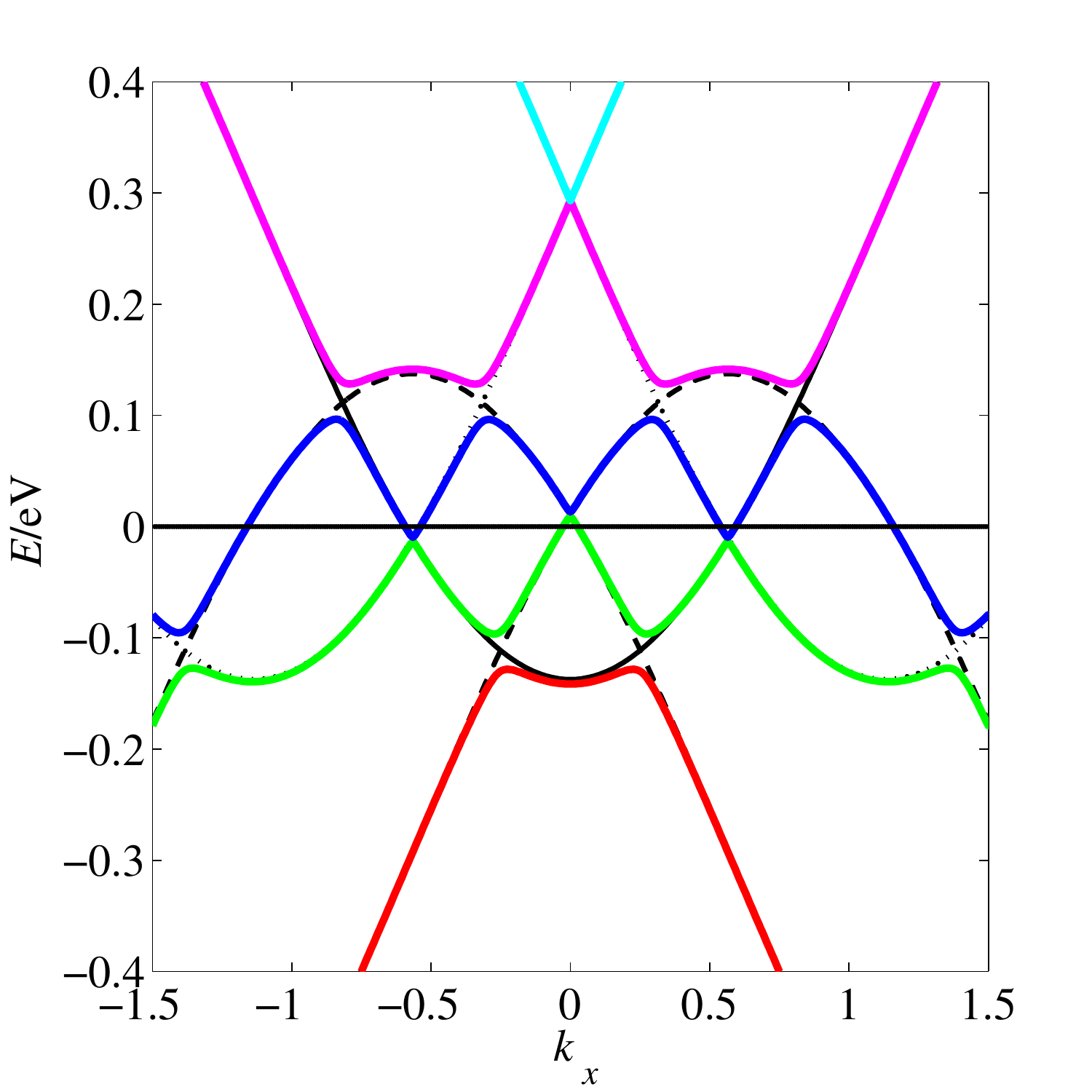}\includegraphics[width=3.25in]{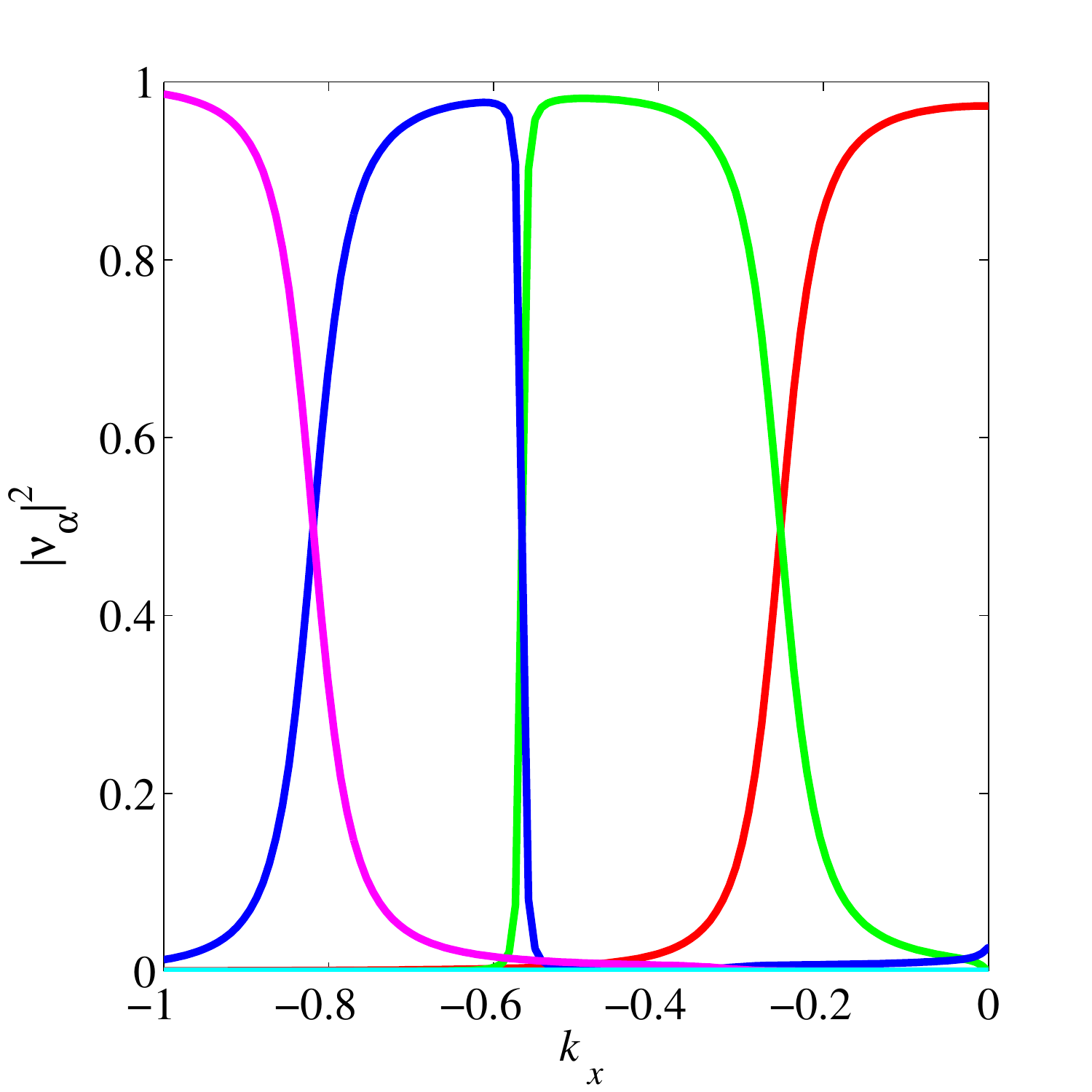}
\caption{Same as Fig.9 except that $\delta k_y$=1.2 }
\end{figure}

\begin{figure}[h]
\includegraphics[width=6.25in]{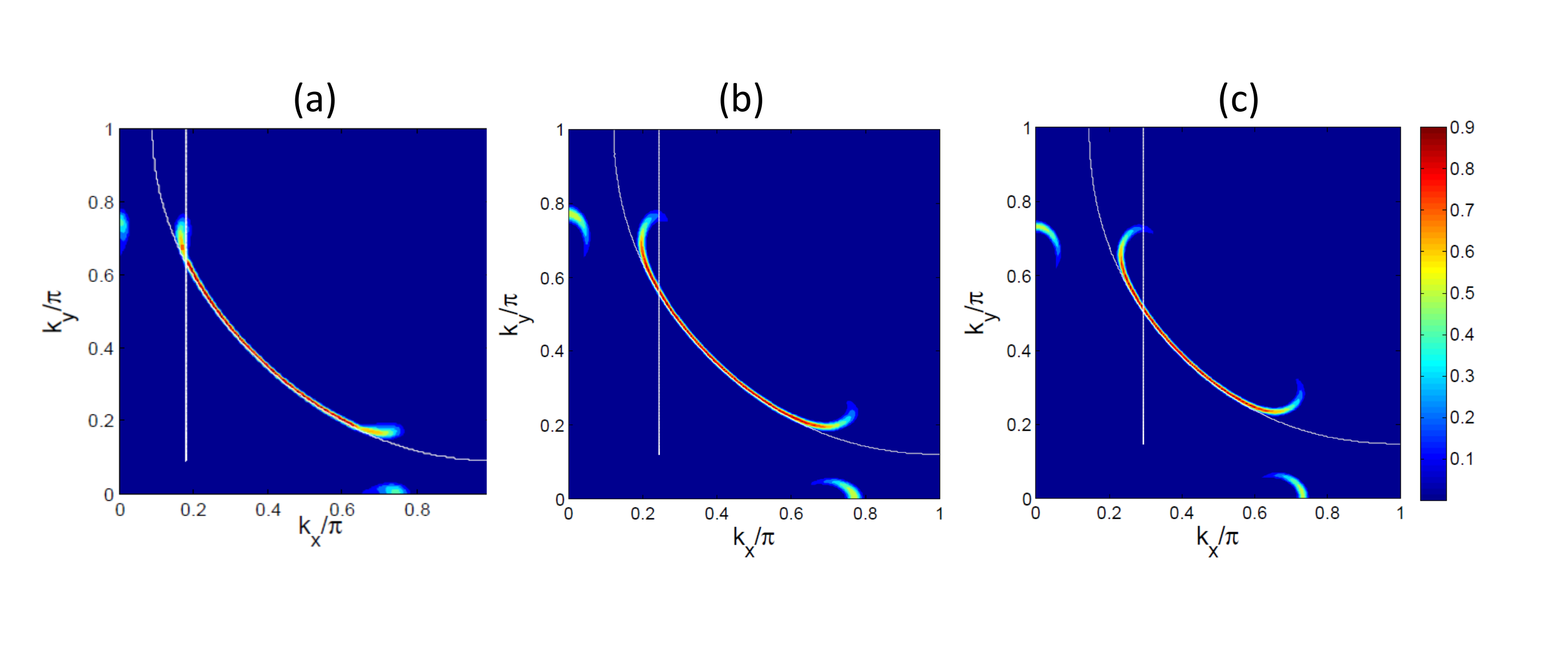}
\caption{Spectral weight at the Fermi level. Same as Fig 4 except that (a)  $\Delta_0=100$~meV, p=0.207. (b) p=0.157 and  $\Delta_0=150$~meV , (c) p=0.112 and  $\Delta_0=250$~meV.}
\end{figure}
 
Note that at the crossing at $k_x=\pm Q$ mentioned  above between the original band and one shifted by 2Q (black and dotted lines in the figures), one combination is decoupled, thereby pinning the blue band to the crossing point. As $k_y$ continues to move away from $\pi$, the crossing point comes down in energy and collides with the top of the green band. This happens near $\delta k_y$=1.2 [Fig.12(a)] and produces a very small gap. Judging from the spectral weight shown in Fig.12(b), the bands are essentially crossing without much coupling and a robust Fermi surface arc is in place beyond this point. The position where the bare band crossing crosses the Fermi level is the point where the vertical white line crosses the bare Fermi surface shown in Fig 4. We emphasize that so far we have set the direct coupling to the CDW to be zero, so that the gap at the tip of the arc is a weak induced gap through the pairing. 
In Fig.13 we show the spectral weight of the Fermi surface for different doping, showing how the Fermi arc shrinks with underdoping.

The energy spectrum for a PDW was calculated by Baruch and Orgad \cite{28b} earlier.  They considered a unidirectional PDW with a wave vector $\delta$ which is not tied to the Fermi surface spanning vector $Q$, but is close to it.  For a sufficiently large gap, the qualitative feature near $(0,\pi)$ is similar to ours.  However, they assume a pairing amplitude which is independent of $k_y$ whereas we assume that it is large only near $(0,\pi)$.  As a result they find a gap structure also near $(\pi,0)$ while we do not if we treated a unidirectional CDW.




\begin{thebibliography}{99}
\bibitem{21b}
Strictly speaking the assumption that the PDW momentum is 2K is justified for weak0 coupling.  For strong coupling we may expect a small deviation.
\bibitem{16a}
H-D Han,O. Vafek, A. Yazdani and S-C Zhang, Phys. Rev. Lett. {\bf 93}, 187003 (2004). This version of PDW has a period which is the same as the CDW and is different from other examples discussed here.
\end{thebibliography}
\end{document}